\documentclass[journal=jctcce,manuscript=article]{achemso}
\setkeys{acs}{articletitle = true}
\usepackage{chemformula}
\usepackage[T1]{fontenc}
\usepackage{blindtext}
\usepackage{amsfonts}
\usepackage{enumitem}
\usepackage{amssymb}
\usepackage{multirow}
\usepackage{xcolor}

\newcommand{\onlinecite}[1]{\hspace{-1 ex} \nocite{#1}\citenum{#1}}
\def\half{\frac{1}{2}}

\author{Micha{\l} Lesiuk}
\email{lesiuk@tiger.chem.uw.edu.pl}
\affiliation{
Faculty of Chemistry, University of Warsaw, \\ Pasteura 1, 02-093 Warsaw, Poland}
\date{\today}

\title[]{Near-exact CCSDT energetics from rank-reduced formalism supplemented by non-iterative 
corrections}


\begin{document}

\begin{tocentry}
\includegraphics[scale=0.80]{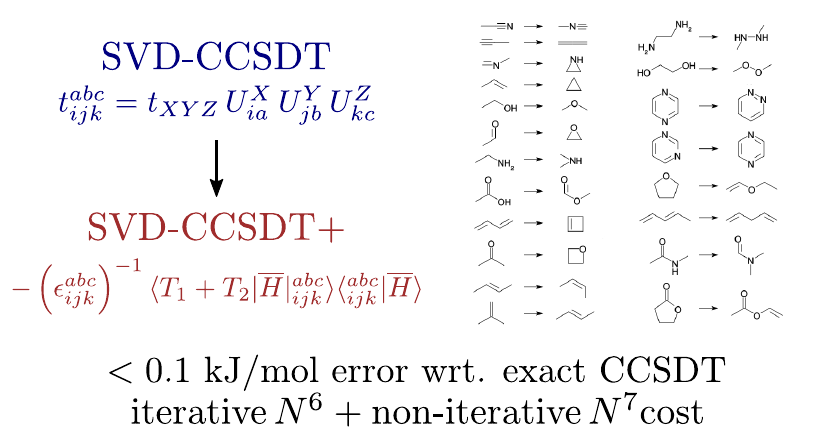}
\end{tocentry}

\renewcommand{\arraystretch}{1.5}

\begin{abstract}
We introduce a non-iterative energy correction, added on top of the rank-reduced coupled-cluster 
method with single, double, and triple substitutions, that accounts for excitations 
excluded from the parent triple excitation subspace. The formula for the correction is
derived by employing the coupled-cluster Lagrangian formalism with an additional assumption that 
the parent excitation subspace is closed under the action of the Fock operator. Owning to the 
rank-reduced form of the triple excitation amplitudes tensor, the computational cost of evaluating 
the correction scales as $N^7$ with the system size, $N$. The accuracy and computational efficiency 
of the proposed method is assessed both for total and relative correlation energies. We show that the non-iterative 
correction can fulfill two separate 
roles. If an accuracy level of a fraction of kJ/mol is sufficient for a given system the correction 
significantly reduces the dimension of the parent triple excitation subspace needed in the iterative 
part of the calculations. Simultaneously, it enables to reproduce the exact CCSDT results to an 
accuracy level below 0.1 kJ/mol with a larger, yet still reasonable, dimension of the parent 
excitation subspace. This typically can be achieved at a computational cost only several times 
larger than required for the CCSD(T) method. The proposed method retains black-box features of the 
single-reference coupled-cluster theory; the dimension of the parent excitation subspace remains 
the only additional parameter that has to be specified.
\end{abstract}

\newpage
\section{Introduction}
\label{sec:intro}

Over the past decades coupled-cluster (CC) 
theory\cite{coester58,coester60,cizek66a,cizek66b,cizek71,cizek72} has been propelled into 
the 
position of one of the most important electronic structure methods. This success can be largely 
attributed to the 
rigorous size-extensivity as well as rapid convergence towards the full configuration-interaction 
limit with the excitation level, see Refs.~\onlinecite{crawford00,bartlett07} for recent reviews. 
However, these virtues come at a price of a rather steep 
scaling of the computational costs of the calculations with the system size. Therefore, many 
approaches such as optimized virtual orbital space\cite{adam87,adam88,neo05,pitoniak06}, frozen 
natural orbitals\cite{sosa89,taube05,taube08,deprince13}, orbital-specific virtuals 
\cite{yang11,kura12,yang12,schutz13}, and local correlation treatments based on 
local pair natural orbitals\cite{neese09b,riplinger13a,riplinger13b,liakos15,schwilk17} were 
proposed in the literature to alleviate this problem. Recently, a new idea to reduce the cost 
of CC (and related) methods has emerged which draws inspiration from the field of 
applied mathematics~\cite{kolda09} and employs tensor decomposition techniques to the wavefunction 
parameters~\cite{scuseria08,bell10,kinoshita03,hino04,benedikt11,benedikt13,schutski17,
mayhall17,parrish19, hohenstein19,lesiuk19,lesiuk20}.
These techniques rely on representing the information contained in the cluster excitation 
amplitudes by a combination of lower-rank quantities. While the basic idea is simple enough, the 
real difficulty lies in selecting a suitable 
excitation subspace and then solving the CC equations within this subspace without 
``unpacking'' the compressed quantities to their initial rank at any stage of the calculations. In 
parallel, similar ideas have also been applied to compression of electron repulsion integrals (ERI) 
leading to the development of the tensor hypercontraction (THC) 
format~\cite{hohenstein12,parrish12} and its efficient 
implementations~\cite{parrish13b,schumacher15,lu15,lee20,matthews20} which enabled to reduce the 
scaling of various electronic structure 
methods~\cite{hohenstein12b,hohenstein13a,hohenstein13b,shenvi13,shenvi14,parrish14,lu17,song16,
song17, song18}.

Recently we have reported~\cite{lesiuk20} application of the Tucker-3 
decomposition~\cite{tucker66,delath00} to the full CCSDT 
theory~\cite{noga87,scuseria88} with the CC triply-excited amplitudes tensor 
represented as (details of the 
notation 
are given further in the text)
\begin{align}
\label{tuck1}
&t_{ijk}^{abc} \approx t_{XYZ} \,U^X_{ai}\,U^Y_{bj} \,U^Z_{ck}.
\end{align}
The expansion tensors $U^X_{ai}$ are obtained upfront by higher-order singular-value 
decomposition (SVD) 
of approximate amplitudes, and the quantity $t_{XYZ}$ is the compressed amplitude tensor. The 
method 
based on Eq. (\ref{tuck1}) shall be referred to as SVD-CCSDT in the remainder of the text. The main 
advantage of the 
decomposition format (\ref{tuck1}) is that the effective dimension of the $t_{XYZ}$ tensor that is
sufficient to maintain a constant relative accuracy in the correlation energy grows only linearly 
with the system size, $N$. Without the compression, i.e., if all possible excitations were 
included in Eq. (\ref{tuck1}), this dimension would grow quadratically. This reduction solves 
two important problems associated with the application of the CCSDT theory to larger systems. 
First, the memory storage requirements are reduced from being proportional to $N^6$ to the level of 
$N^4$ because the full-rank amplitudes $t_{ijk}^{abc}$ are never explicitly formed and only their 
compressed counterparts ($t_{XYZ}$) are stored. Second, by careful 
factorization of the CC equations and manipulating the order of tensor contractions 
one can reduce the scaling of the computational costs from $N^8$ (characterizing for the 
uncompressed CCSDT method) down to $N^6$.

In this work we expand upon the rank-reduced CCSDT theory introduced in 
Ref.~\onlinecite{lesiuk20}. We 
propose a non-iterative, 
i.e., single-step, energy correction added on top of the SVD-CCSDT 
result. The purpose of this correction is to reduce the error with respect to the exact 
(uncompressed) CCSDT method by approximately 
accounting for triple excitations absent in the parent SVD subspace. The idea of adding a 
non-iterative 
correction to a converged coupled cluster result in order to account for, e.g., higher 
excitations excluded from the iterative model, is not new. In fact, numerous 
approaches have 
been proposed in the literature to derive such corrections. Historically, the first developments of 
this type were guided by the ordinary M\o{}ller-Plesset perturbation theory~\cite{moller34} where 
the 
Hartree-Fock determinant serves as the zeroth-order wavefunction. However, this approach turned 
out to be suboptimal as best exemplified by the success of the CCSD(T) 
theory~\cite{ragha89} over the earlier CCSD[T] method~\cite{noga87b}. The latter is based solely on 
the usual perturbative arguments 
while the 
former 
includes, seemingly arbitrarily, a single higher-order term (out of many possible). A justification 
of this choice was presented by Stanton~\cite{stanton97} who treated CCSD as the zeroth-order state 
and employed a 
formalism rooted in the equation-of-motion (EOM) theory~\cite{stanton93} to derive the 
non-iterative correction.  
With the help of L\"{o}wdin's partitioning technique of the EOM Hamiltonian one can then show that 
the troubling higher-order term appears naturally and should be treated on an equal footing. The 
EOM-like formalism was further developed and refined by Gwaltney and  
Head-Gordon~\cite{gwaltney00,gwaltney01} who derived a complete second-order correction to the CCSD 
energy, termed CCSD(2). 
Subsequent work in this field by Hirata and collaborators~\cite{hirata01,hirata04,shiozaki07} 
culminated in the introduction of the  
CC($m$)PT($n$) systematic hierarchy of methods. Sometime later the lack of order-by-order 
size consistency of the EOM-like approaches was recognized~\cite{eriksen14a}. Eriksen \emph{et 
al.}~\cite{eriksen14b} showed how this 
problem can be avoided if non-iterative corrections are derived by expanding the CC 
Langrangian of a higher-order method around a lower-order one. This leads to a hierarchy of methods 
such as CCSD(T-$n$) and CCSDT(Q-$n$) which are rigorously size-extensive in each order $n$ 
separately, not only in their limit. A similar Lagrangian-based formulation was constructed by 
Kristiansen \emph{et al.}~\cite{kristiansen16} and showed an improved convergence characteristics. 
The size-consistency problems are also avoided in the framework of cluster perturbation theory 
developed by Paw{\l}owski and 
collaborators~\cite{pawlowski19a,pawlowski19b,pawlowski19c,pawlowski19d,pawlowski19e}.
It is also important to point out the papers of Piecuch and 
collaborators~\cite{kowalski00,piecuch02,piecuch04,piecuch05} who derived non-iterative corrections 
to the CC methods employing the so-called 
method-of-moments CC theory. This methodology has been progressively refined over the years, with 
recent introduction of an impressively general CC($P$;$Q$) hierarchy~\cite{shen12a,shen12b,shen12c}.

While the body of work published in the literature that deals with derivation of non-iterative 
corrections to 
the CC energies is large, none of the available formulas can be applied straight away in the 
SVD-CCSDT context without encountering serious problems of either theoretical or practical nature. 
These 
difficulties stem from the fact that in the SVD-CCSDT theory the parent triples excitation 
subspace is not spanned by some set of individual excitations. Instead, the basis of this subspace 
is composed of linear combinations of triple excitations which are found automatically by a 
procedure described 
in Ref.~\onlinecite{lesiuk19}. To accommodate this problem we introduce a partitioning of the 
triple excitation space and generalize the Langrangian formalism of Eriksen \emph{et 
al.}~\cite{eriksen14a} This 
leads to a formula for a non-iterative energy correction that is 
similar in nature to the celebrated CCSD(T) method and, critically, can also be evaluated with the 
computational 
cost proportional to $N^7$.

The practical reason for introducing the non-iterative correction is twofold. First, it has been 
shown in Ref.~\onlinecite{lesiuk20} that the practical accuracy limit of the SVD-CCSDT method is, 
on average, a 
fraction of kJ/mol in relative energies. Provided that this level of accuracy is 
sufficient 
for the task at hand, the non-iterative correction enables a considerable reduction of the 
dimension of 
the parent excitation subspace needed in the iterative part of the calculations. This is 
advantageous 
because for small parent subspaces the timing of the SVD-CCSDT calculations is only a small 
multiple 
of the CCSD calculations for the same system. Simultaneously, we show that the inclusion of the 
non-iterative correction allows to reduce the error with respect to the uncompressed CCSDT by 
roughly an order of magnitude. Therefore, if the SVD subspace is large enough, accuracy levels
below 0.1 kJ/mol become accessible. One can expect this accuracy to be sufficient in all 
but the most accurate studies concerning polyatomic molecules. In fact, other sources of error, 
such basis set incompleteness, are typically of the same magnitude or larger.

\section{Theory}
\label{sec:theory}

\subsection{Preliminaries}
\label{subsec:pre}

For the convenience of the readers we begin by defining the notation that is adopted throughout 
the 
present paper and provide a short outline of the SVD-CCSDT theory. We employ 
the canonical Hartree-Fock (HF) determinant, denoted $|\phi_0\rangle$, as the reference 
wavefunction. The orbitals that are occupied in the reference are denoted by the symbols $i$, $j$, 
$k$, etc., and the unoccupied (virtual) orbitals by the symbols $a$, $b$, $c$, etc. When the 
occupation of the orbital is not specified general indices $p$, $q$, $r$, etc. are employed. The 
number of occupied and virtual orbitals in a given system is written as $O$ and $V$, respectively. 
The HF 
orbital energies are denoted by $\epsilon_p$. For further use we also introduce the following 
conventions: $\langle 
A\rangle \stackrel{\mbox{\tiny def}}{=} \langle \phi_0 | A \phi_0 \rangle$ and $\langle 
A|B\rangle \stackrel{\mbox{\tiny def}}{=} \langle A \phi_0|B \phi_0 \rangle$ for arbitrary operators 
$A$, $B$. The Einstein convention for summation over repeated indices is employed unless explicitly 
stated otherwise. The standard partitioning of the electronic Hamiltonian, $H=F+W$, into the sum of 
the Fock operator ($F$) and the fluctuation potential ($W$) is adopted throughout the 
paper.

The SVD-CCSDT method is a variant of the CC theory and employs the exponential 
parametrization of the electronic wavefunction, $|\Psi\rangle = e^{T_{\rm SVD}}\,|\phi_0\rangle$. 
The cluster operator $T_{\rm SVD}$ contains single, double, and triple excitation operators, 
$T_{\rm SVD}=T_1+T_2+T_3^{\rm SVD}$. The singly and doubly excited components assume the same form 
as in the uncompressed CCSDT theory
\begin{align}
\label{t12}
 T_1 = t_i^a\,E_{ai}, \;\;\;
 T_2 = \frac{1}{2}\,t_{ij}^{ab} \,E_{ai}\,E_{bj},
\end{align}
where $t_i^a$, $t_{ij}^{ab}$ are the cluster amplitudes, and $E_{pq}=p^\dagger_\alpha q_\alpha + 
p^\dagger_\beta q_\beta$ are the spin-adapted singlet orbital replacement 
operators~\cite{paldus88}. The triply 
excited amplitudes $t_{ijk}^{abc}$ are subject to the Tucker-3 compression, see 
Eq. (\ref{tuck1}), and hence the triple excitation operator reads
\begin{align}
\label{t3}
 T_3^{\rm SVD} = \frac{1}{6}\, t_{ijk}^{abc}\,E_{ai}\,E_{bj}\,E_{ck} = 
 \frac{1}{6}\,t_{XYZ}\,U^X\,U^Y\,U^Z,
\end{align}
where $U^X=U^X_{ai}\,E_{ai}$. Throughout this paper the symbols $X$, $Y$, $Z$ 
are employed to denote the elements of the parent subspace (SVD subspace) of the triply 
excited amplitudes. 
The 
length of the expansion in Eq. (\ref{tuck1}) is denoted by $N_{\mathrm{SVD}}$. For further use 
we also define the symbols $|_i^a\rangle=E_{ai}|\phi_0\rangle$, 
$|_{ij}^{ab}\rangle=E_{ai}E_{bj}|\phi_0\rangle$, etc., to denote excited-state configurations and 
similarly
\begin{align}
\label{psixyz}
 |XYZ\rangle=U^X_{ai}\,U^Y_{bj} \,U^Z_{ck}\,|_{ijk}^{abc}\rangle=U^X\,U^Y\,U^Z 
 |\phi_0\rangle.
\end{align}
The amplitudes $t_i^a$, $t_{ij}^{ab}$, and $t_{XYZ}$ are found by solving the SVD-CCSDT 
equations obtained by projecting $e^{-T_{\rm SVD}} H e^{T_{\rm SVD}} |\phi_0\rangle=0$ onto the 
proper subset of excited configurations. The expansion vectors $U^X_{ai}$ are obtained upfront by 
singular-value 
decomposition of an approximate triples amplitude tensor
\begin{align}
\label{t32}
 \,^{(2)}t_{ijk}^{abc} = (\epsilon_{ijk}^{abc})^{-1} \langle \,_{ijk}^{abc} | \big[ \widetilde{W}, 
T_2 \big]\rangle,
\end{align}
rewritten as a $OV\times O^2V^2$ rectangular matrix. The symbol $\epsilon_{ijk}^{abc} 
=\epsilon_i+\epsilon_j+\epsilon_k-\epsilon_a-\epsilon_b-\epsilon_c$ stands for 
the three-particle energy denominator and $\widetilde{W}=e^{-T_1}We^{T_1}$ is the $T_1$-transformed 
fluctuation potential. To obtain the optimal 
compression of the full tensor $t_{ijk}^{abc}$ to a desired size ($N_{\mathrm{SVD}}$) one has to 
retain only those vectors $U_{ai}^X$ that correspond to the largest singular values.

In the previous papers devoted to the rank-reduced CC methods including triple excitations, the SVD 
of Eq. (\ref{t32}) was computed using an iterative bidiagonalization method. While this method is 
completely general, it scales as $N^7$ with the system size, more steeply than $N^6$ cost of the 
SVD-CCSDT iterations. While it has been shown~\cite{lesiuk20} that the prefactor of the former 
procedure is small and finding the SVD subspace does not constitute a bottleneck at present, this 
may change for larger molecules. To avoid this problem in this work we propose an alternative 
algorithm for determination of the parent triple excitations subspace from the approximate 
amplitudes (\ref{t32}). A detailed derivation of the method is included in the Supporting 
Information, along with numerical examples that confirm its reliability. The new algorithm gives 
exactly same results as its predecessor (assuming exact arithmetic), but it possesses a rigorous 
$N^6$ scaling of the computational costs with the system size which is advantageous in applications 
to larger systems. Moreover, the new method is non-iterative in nature which eliminates possible 
accumulation of numerical noise and convergence problems one may encounter in iterative schemes. The 
proposed method is used by default in all SVD-CCSDT calculations reported further in the paper.

To decompose the four-index electron repulsion integrals tensor, $(pq|rs)$, we employ the 
robust density-fitting 
approximation~\cite{whitten73,baerends73,dunlap79,alsenoy88,vahtras93} (in the Coulomb 
metric)
\begin{align}
\label{dfint}
 (pq|rs) \approx B_{pq}^Q\,B_{rs}^Q,\;\;\;\mbox{with}\;\;\;B_{pq}^Q = 
(pq|P)\,[\mathbf{V}^{-1/2}]_{PQ},
\end{align}
where $(pq|P)$ and $V_{PQ}=(P|Q)$ are the three-center and two-center electron repulsion integrals, 
respectively, as defined in Ref.~\onlinecite{katouda09}. The capital letters $P$, $Q$ are employed 
throughout the present 
work to denote the elements of the auxiliary basis set (ABS). The formula (\ref{dfint}) preserves 
all physical symmetries and positive-definiteness of the initial electron repulsion integrals 
tensor~\cite{wirz17}. The 
number of ABS functions is denoted $N_{\mathrm{aux}}$ further in the text and it scales linearly 
with the size of the system. The same formula is applied also for the $T_1$-transformed 
two-electron 
integrals~\cite{koch94}, $(pq\widetilde{|}rs)\approx B_{\widetilde{pq}}^Q\,B_{\widetilde{rs}}^Q$, 
that correspond 
to the similarity-transformed Hamiltonian $\widetilde{H}=e^{-T_1}He^{T_1}$. The only difference is 
that the three-center integrals $B_{\widetilde{pq}}^Q$ are given in the $T_1$-transformed, rather 
than the canonical, orbital basis. Let us also point out that all equations derived in the present 
work 
remain valid also for the Cholesky decomposition~\cite{beebe77,koch03,pedersen04,folkestad19} of the 
electron repulsion integrals since both 
methods share the same formal expression, Eq.~(\ref{dfint}).

\subsection{Partitioning of the triple excitation subspace}
\label{subsec:trispace}

The complete space of triple excitations is defined as $\mathrm{span}(\mu_3)$, where 
the symbol $\mathrm{span}(S)$ stands for a linear span of a set of vectors $S$, and $\mu_3$ is the 
following set
\begin{align}
 \mu_3 = \Big\{ |_{ijk}^{abc}\rangle\;|\;a<b\leq c \Big\}.
\end{align}
Note that the set of vectors $\mu_3$ is not linearly independent; for example, among six 
possible functions $|_{ijk}^{abc}\rangle$, $a<b<c$, with a fixed set of indices only five are 
independent~\cite{paldus88} as a consequence of the relation
\begin{align}
 |_{ijk}^{abc}\rangle+|_{ikj}^{abc}\rangle+|_{jik}^{abc}\rangle+|_{kji}^{abc}\rangle
 +|_{kij}^{abc}\rangle+|_{jki}^{abc}\rangle=0.
\end{align}
However, we require only that $\mu_3$ spans the complete space of triple excitations and the fact 
that it is not minimal bears no negative consequences in the present context. Next we introduce a 
set composed of the SVD vectors
\begin{align}
 \mu_3^{\rm SVD} = \Big\{ |XYZ\rangle\;|\;X\leq Y\leq Z \Big\},
\end{align}
where $|XYZ\rangle$ was defined in Eq. (\ref{psixyz}).
The span of this set, $\rm{span}(\mu_3^{\rm SVD})$, is shortly referred to as the parent subspace 
or the SVD subspace. In 
general, the set $\mu_3^{\rm SVD}$ may also contain linearly dependent elements. Finally, the third 
set of vectors, $\mu_3^\bot$, spans the orthogonal complement of $\mu_3^{\rm SVD}$. This means that 
$\mu_3^\bot$ has the following two properties
\begin{align}
\label{prop1}
&\mathrm{span}(\mu_3^{\rm SVD}\cup\mu_3^\bot) = \mathrm{span}(\mu_3),\\
\label{prop2}
&\langle\mu_3^{\rm SVD}|\mu_3^\bot\rangle=0,
\end{align}
where the second equation is understood to hold separately for each pair of vectors from the 
sets $\mu_3^{\rm SVD}$ and $\mu_3^\bot$. A natural way to generate the vectors $\mu_3^\bot$ is to 
project out the set $\mu_3^{\rm SVD}$ from $\mu_3$. This choice is not unique, 
but the final results of this work do not depend on a particular procedure employed to generate 
$\mu_3^\bot$ provided that the relationships (\ref{prop1}) and (\ref{prop2}) are strictly true.

To simplify the subsequent derivations we have to specify how the Fock operator acts within the SVD 
subspace and its orthogonal complement. First, it is convenient to exploit the 
rotational freedom among the quantities $U_{ai}^X$ and enforce the relationship
$U^X_{ai}\, U^Y_{ai}\, (\epsilon_i-\epsilon_a) = \epsilon_X\,\delta_{XY}$,
where $\epsilon_X$ are some real-valued constants. This makes the Fock operator diagonal in the SVD 
subspace in the sense that
\begin{align}
 \langle X'Y'Z'|F|XYZ\rangle=\left( \epsilon_X + \epsilon_Y + \epsilon_Z \right) 
 \delta_{XX'}\,\delta_{YY'}\,\delta_{ZZ'}.
\end{align}
In the present work we assume a stronger condition that the SVD subspace is closed under the 
action of the Fock operator. In other words
\begin{align}
\label{gbc1}
 F|XYZ\rangle = \left( \epsilon_X + \epsilon_Y + \epsilon_Z \right)|XYZ\rangle,
\end{align}
which, in general, constitutes an approximation when the SVD subspace is not complete. An immediate 
consequence of Eq. (\ref{gbc1}) is the 
relationship
\begin{align}
\label{prop3}
&\langle\mu_3^{\rm SVD}|F|\mu_3^\bot\rangle=\langle\mu_3^\bot|F|\mu_3^{\rm SVD}\rangle=0,
\end{align}
which plays important role in the derivations presented in the next section. The accuracy of this 
approximation depends on the size of the SVD subspace and in Sec. \ref{sec:numer} we 
present a numerical verification of Eq. (\ref{prop3}) for realistic systems. 

For consistency, we additionally introduce sets $ \mu_1 = \Big\{ |_{i}^{a}\rangle \Big\}$ and 
$ \mu_2 = \Big\{ |_{ij}^{ab}\rangle\;|\;a\leq b \Big\}$, so that $\rm{span}(\mu_1)$ and 
$\rm{span}(\mu_2)$ are complete spaces of single and double excitations, respectively.

\subsection{Perturbative correction to the SVD-CCSDT energy}
\label{subsec:pert}

In this subsection we derive a non-iterative correction that approximately accounts for 
triple excitations excluded from the parent SVD subspace. First, we introduce the SVD-CCSDT 
Lagrangian~\cite{fritz86,salter89,jorgensen88,helga89,koch90,koch97} and briefly discuss its most 
salient features. Next, we follow the idea of Eriksen \emph{et al.}~\cite{eriksen14a} and expand 
the Lagrangian of the exact CCSDT method around the SVD-CCSDT Lagrangian; the difference between 
them defines the desired energy correction. Finally, a perturbative scheme is introduced which 
allows to extract the components of the energy correction that are of the leading order in the 
fluctuation potential.

Let us begin by defining the Lagrangian~\cite{fritz86,salter89,jorgensen88,helga89,koch90,koch97} 
of the SVD-CCSDT method. To this end we introduce the second excitation operator $\mathcal{L}_{\rm 
SVD} = 
1+\mathcal{L}_1+\mathcal{L}_2+\mathcal{L}_3^{\rm SVD}$ which is fully analogous to the usual 
cluster operator $T_{\rm SVD}$ defined by Eqs. (\ref{t12}) and (\ref{t3}), but contains the 
additional unity term\footnote{Note that the operator $\mathcal{L}_{\rm SVD}$ is 
usually written in the literature~\cite{bartlett07} as $\mathcal{L}_{\rm SVD}=1+\Lambda^\dagger$, 
where $\Lambda$ is a pure de-excitation operator, but this notation is not particularly convenient 
from the point of view of the present work.}. For simplicity, the triply excited component 
$\mathcal{L}_3^{\rm SVD}$ is expanded in the same SVD subspace as $T_3^{\rm SVD}$. For brevity, 
further in the text we refer to the amplitudes of $\mathcal{L}_{\rm SVD}$ as ``multipliers''.

Under these provisions we can write down the SVD-CCSDT Lagrangian in full form
\begin{align}
\label{svdlag}
 L_{\rm SVD} = \langle \mathcal{L}_{\rm SVD} | e^{-T_{\rm SVD}}He^{T_{\rm SVD}} \rangle.
\end{align}
By recalling the SVD-CCSDT amplitude equations~\cite{lesiuk20}
\begin{align}
\label{svdteq12}
 &\langle \mu_n | e^{-T_{\rm SVD}}He^{T_{\rm SVD}} \rangle = 0,\;\;\; n=1,2, \\
 \label{svdteq3}
 &\langle \mu_3^{\rm SVD} | e^{-T_{\rm SVD}}He^{T_{\rm SVD}} \rangle = 0,
\end{align}
one can show the value of $L_{\rm SVD}$ is equal to the SVD-CCSDT correlation energy for
converged $T_{\rm SVD}$ amplitudes. Additionally, by construction $L_{\rm SVD}$ is variational with 
respect to the $\mathcal{L}_{\rm SVD}$ amplitudes. However, in contrast to the usual CC energy 
formula, we require that this quantity is variational also with respect to the 
cluster amplitudes $T_{\rm SVD}$. Minimization over $T_{\rm SVD}$ gives 
the following set of linear equations for the multipliers
\begin{align}
\label{lambda1}
 &\langle \mathcal{L}_{\rm SVD} | \Big[ \overline{H}, \mu_n \Big] \rangle = 0,\;\;\; n=1,2, \\
 \label{lambda2}
 &\langle \mathcal{L}_{\rm SVD} | \Big[ \overline{H}, \mu_3^{\rm SVD} \Big] \rangle = 0,
\end{align}
where we have introduced a shorthand notation $\overline{H}=e^{-T_{\rm SVD}}He^{T_{\rm SVD}}$ for 
the similarity-transformed SVD-CCSDT Hamiltonian. Completely analogous definitions of the 
Lagrangian hold for the exact 
(uncompressed) CCSDT theory. By denoting the exact CCSDT amplitudes by $T$ and the auxiliary 
operator by $\mathcal{L}$, the CCSDT Lagrangian is written as $L_T = \langle \mathcal{L}| 
e^{-T}He^{T} \rangle$. The equations defining $\mathcal{L}$ are again found by requiring that 
the Lagrangian is variational with respect to the cluster amplitudes.

Our first goal is to parametrize the exact CCSDT Lagrangian ($L_T$) 
around the SVD-CCSDT Lagrangian ($L_{\rm SVD}$) following the idea of Eriksen \emph{et 
al.}~\cite{eriksen14a} To 
this end we rewrite the CCSDT cluster operators as $T=T_{\rm SVD}+\delta T$ and 
$\mathcal{L} = 
\mathcal{L}_{\rm SVD} + \delta \mathcal{L}$, where $\delta T$ and $\delta \mathcal{L}$ are the 
correction terms that have to be determined. They can be further expanded as
\begin{align}
\label{deltat}
 \delta T = \delta T_1 + \delta T_2 + \delta T_3^{\rm SVD} + \delta T_3^\bot,
\end{align}
and analogously for the second quantity $\delta \mathcal{L}$. As the notation suggests, the 
components $\delta T_3^{\rm SVD}$ and $\delta T_3^\bot$ include excitations only to the 
configurations 
belonging to $\mu_3^{\rm SVD}$ and $\mu_3^\bot$, respectively. Because the two subspaces are 
orthogonal there is no double-counting of the excitations; moreover, the sum of both operators 
covers 
all possible excitations in the system. The division introduced in Eq. (\ref{deltat}) also has a 
clear 
interpretation -- the operators $\delta T_1$, $\delta T_2$, and $\delta T_3^{\rm SVD}$ are 
responsible for ``relaxation'' of the excitation amplitudes that are already included in the 
SVD-CCSDT 
wavefunction, while the operator $T_3^\bot$ corrects for the excitations outside the SVD subspace.

Let us to rewrite the exact CCSDT Lagrangian as
\begin{align}
 L_T = \langle \mathcal{L}_{\rm SVD} + \delta \mathcal{L}| e^{-\delta T}\,\overline{H} e^{\delta T}
\rangle,
\end{align}
and employ the nested commutator expansion to eliminate the operator 
exponentials. After some rearrangements one obtains
\begin{align}
\label{step1}
\begin{split}
 L_T &= \langle \mathcal{L}_{\rm SVD} | \overline{H} \rangle + \langle \delta \mathcal{L} | 
\overline{H} \rangle + \langle \mathcal{L}_{\rm SVD} | \Big[ \overline{H}, \delta T \Big] \rangle \\
&+ \sum_{n=2}\frac{1}{n!}\,\langle\mathcal{L}_{\rm SVD}|\Big[ 
\overline{H}, \delta T \Big]_n \rangle
+ \sum_{n=1}\frac{1}{n!}\,\langle\delta \mathcal{L}|\Big[ 
\overline{H}, \delta T \Big]_n \rangle
\end{split}
\end{align}
where $[X,Y]_n$ is a shorthand notation for $n$-tuply nested commutators, i.e. $[X,Y]_1=[X,Y]$ and 
$[X,Y]_n=\big[[X,Y]_{n-1},Y\big]$ for $n\geq2$. We immediately recognize the first term as the 
SVD-CCSDT Lagrangian, $L_{\rm SVD} =\langle\mathcal{L}_{\rm SVD} | \overline{H} \rangle$, cf. Eq. 
(\ref{svdlag}). Therefore, further in the text we consider only the difference $\delta E = L_T - 
L_{\rm SVD}$ which is the desired energy correction. Next, 
we observe that $\langle \delta \mathcal{L} |\overline{H} \rangle = \langle \delta 
\mathcal{L}_3^\bot |\overline{H} \rangle$, because projections on the other components of 
$\mathcal{L}$, such as $\langle \mathcal{L}_2 |\overline{H} \rangle$, vanish due to the SVD-CCSDT 
stationarity conditions, Eqs. (\ref{svdteq12}) and (\ref{svdteq3}). Next, the third term in the 
above equation can be rewritten as $\langle 
\mathcal{L}_{\rm SVD} | \Big[ \overline{H}, \delta T \Big] \rangle = \langle \mathcal{L}_{\rm SVD} | 
\Big[ \overline{H}, \delta T_3^\bot \Big] \rangle$, because the remaining contributions are zero by 
the virtue of Eqs. (\ref{lambda1}) and (\ref{lambda2}). The last two terms in Eq. (\ref{step1}) 
remain unaltered at this point. We have arrived at the following 
formula for the energy correction
\begin{align}
\label{interm1}
\begin{split}
 \delta E = \langle \delta \mathcal{L}_3^\bot | \overline{H} \rangle 
 + \langle \mathcal{L}_{\rm SVD} | \Big[ \overline{H}, \delta T_3^\bot \Big] \rangle
 + \sum_{n=2}\frac{1}{n!}\,\langle\mathcal{L}_{\rm SVD}|\Big[ 
\overline{H}, \delta T \Big]_n \rangle
+ \sum_{n=1}\frac{1}{n!}\,\langle\delta \mathcal{L}|\Big[ 
\overline{H}, \delta T \Big]_n \rangle.
\end{split}
\end{align}
To proceed further we split the similarity-transformed Hamiltonian into two contributions, 
$\overline{H} = \overline{F} + \overline{W}$, where $\overline{F} = e^{-T_{\rm SVD}}Fe^{T_{\rm 
SVD}}$ and $\overline{W} = e^{-T_{\rm SVD}}We^{T_{\rm SVD}}$. Note that since $F$ is a one-electron 
operator we have $\overline{F}=F+\big[F,T_{\rm SVD}\big]$, i.e. the multiply nested commutators 
vanish. Moreover, for any purely excitation operator $X$ the relationship $\big[ \overline{F}, 
X\big]=\big[ F, X\big]$ holds. This allows to rewrite Eq. (\ref{interm1}) as
\begin{align}
\begin{split}
 \delta E &= \langle \delta \mathcal{L}_3^\bot | \overline{F} \rangle 
 + \langle \delta \mathcal{L}_3^\bot | \overline{W} \rangle 
 + \langle \mathcal{L}_{\rm SVD} | \Big[ F, \delta T_3^\bot \Big] \rangle
 + \langle \mathcal{L}_{\rm SVD} | \Big[ \overline{W}, \delta T_3^\bot \Big] \rangle \\
 &+ \langle\delta \mathcal{L}|\Big[ F, \delta T \Big] \rangle +
 \sum_{n=2} \frac{1}{n!}\,\langle\mathcal{L}_{\rm SVD}|\Big[ \overline{W}, \delta T \Big]_n\rangle
 + \sum_{n=1} \frac{1}{n!}\,\langle\delta \mathcal{L}|\Big[ \overline{W}, \delta T \Big]_n \rangle,
\end{split}
\end{align}
where all other terms involving $\overline{F}$ vanished due to an inadequate excitation level.
Up to this point we have introduced no approximations into this formalism. However, to simplify the 
equations further we invoke the condition (\ref{prop3}) and set $\langle\mu_3^{\rm 
SVD}|F|\mu_3^\bot\rangle=\langle\mu_3^\bot|F|\mu_3^{\rm SVD}\rangle=0$ whenever applicable. As 
discussed in Sec. \ref{subsec:trispace} this is an approximation unless the SVD subspace is 
complete. By employing Eq. (\ref{prop3}) one eliminates the first term of the 
above formula, because $\langle \delta \mathcal{L}_3^\bot | \overline{F} \rangle = \langle \delta 
\mathcal{L}_3^\bot | \big[F,T_{\rm SVD}\big]\rangle$, and the third term for the same reason. 
Therefore, we are left with
\begin{align}
\label{lag1}
\begin{split}
 \delta E &= \langle\delta \mathcal{L}|\Big[ F, \delta T \Big] \rangle 
 + \langle \delta \mathcal{L}_3^\bot | \overline{W} \rangle 
 + \langle \mathcal{L}_{\rm SVD} | \Big[ \overline{W}, \delta T_3^\bot \Big] \rangle \\
 &+ \sum_{n=2} \frac{1}{n!}\,\langle\mathcal{L}_{\rm SVD}|\Big[ \overline{W}, \delta T \Big]_n 
\rangle
 + \sum_{n=1} \frac{1}{n!}\,\langle\delta \mathcal{L}|\Big[ \overline{W}, \delta T \Big]_n \rangle.
\end{split}
\end{align}
By construction, the quantity given by Eq. (\ref{lag1}) is variational both with respect to the 
cluster amplitudes and the multipliers. Therefore, a suitable set of equations for the perturbed 
amplitudes in $\delta 
\mathcal{L}$ and $\delta T$ can be obtained by minimization. By differentiating Eq. (\ref{lag1}) 
over the perturbed multipliers ($\delta\mathcal{L}$) and equating the result to zero one obtains 
formulas for all components of the perturbed cluster amplitudes
\begin{subequations}
\begin{align}
\begin{split}
\label{delta12}
 \langle\mu_m | \Big[ F, \delta T_m \Big] \rangle +
 \sum_{n=1} \frac{1}{n!}\,\langle\mu_m|\Big[ \overline{W}, \delta T \Big]_n \rangle = 0,
 \;\;\; m=1,2,
\end{split}\\
\begin{split}
\label{delta3svd}
 \langle\mu_3^{\rm SVD} | \Big[ F, \delta T_3^{\rm SVD} \Big] \rangle +
 \sum_{n=1} \frac{1}{n!}\,\langle\mu_3^{\rm SVD}|\Big[ \overline{W}, \delta T \Big]_n \rangle = 0,
\end{split}\\
\begin{split}
\label{delta3bot}
\langle\mu_3^\bot | \Big[ F, \delta T_3^\bot \Big] \rangle + \langle \mu_3^\bot | 
\overline{W} \rangle +
 \sum_{n=1} \frac{1}{n!}\,\langle\mu_3^\bot|\Big[ \overline{W}, \delta T \Big]_n \rangle = 0,
\end{split}
\end{align}
\end{subequations}
Similarly, by minimization with respect to the cluster amplitudes ($\delta T$) one determines the 
perturbed multipliers
\begin{subequations}
\begin{align}
\begin{split}
 \label{mult12}
 \langle\delta \mathcal{L}_m|\Big[ F, \mu_m \Big] \rangle +
 \langle\delta \mathcal{L}|\Big[ \overline{W}, \mu_m \Big] \rangle + 
 \sum_{n=1} \frac{1}{n!}\,\langle\mathcal{L}_{\rm SVD}+\delta\mathcal{L}|
 \Big[\big[ \overline{W},\delta T \big]_n, \mu_m\Big]\rangle = 0,
\end{split}\\
\begin{split}
 \label{mult3svd}
 \langle\delta \mathcal{L}_3^{\rm SVD}|\Big[ F, \mu_3^{\rm SVD} \Big] \rangle +
 \langle\delta \mathcal{L}|\Big[ \overline{W}, \mu_3^{\rm SVD} \Big] \rangle + 
 \sum_{n=1} \frac{1}{n!}\,\langle\mathcal{L}_{\rm SVD}+\delta\mathcal{L}|
 \Big[\big[ \overline{W},\delta T \big]_n, \mu_3^{\rm SVD}\Big]\rangle = 0,
\end{split}\\
\begin{split}
 \label{mult3bot}
 \langle\delta \mathcal{L}_3^\bot|\Big[ F, \mu_3^\bot \Big] \rangle +
 \langle\mathcal{L}_{\rm SVD}|\Big[ \overline{W}, \mu_3^\bot \Big] \rangle + 
 \sum_{n=1} \frac{1}{n!}\,\langle\mathcal{L}_{\rm SVD}+\delta\mathcal{L}|
 \Big[\big[ \overline{W},\delta T \big]_n, \mu_3^\bot\Big]\rangle = 0.
\end{split}
\end{align} 
\end{subequations}
The formalism given above is not yet practically useful due to the 
computational cost being roughly the same as of the exact CCSDT theory. To eliminate this problem 
we set up a perturbative expansion of the above equations, treating the similarity-transformed Fock 
operator ($\overline{F}$) as the zeroth-order quantity and the similarity-transformed fluctuation 
potential ($\overline{W}$) as the first-order quantity. This leads to the expansion of 
the cluster amplitudes and the multipliers in the orders of the fluctuation potential
\begin{align}
\label{formalt}
 \delta T           = \delta T(0) + \delta T(1) + \delta T(2) + \ldots \\
\label{formall}
 \delta \mathcal{L} = \delta \mathcal{L}(0) + \delta \mathcal{L}(1) + \delta \mathcal{L}(2) + \ldots
\end{align}
where the order of a given term is indicated in the parentheses, e.g. $\delta T(n)$ is the $n$-th 
order component of $\delta T$. Similarly, the energy correction $\delta E$ is also rewritten as a 
sum $\delta E=\delta E(0)+\delta E(1)+\delta E(2)+\ldots$ At each perturbation order the cluster 
operators and multipliers are further split into the components corresponding different excitation 
manifolds, for example, $\delta T(n) = \delta T_1(n) + \delta T_2(n) + \delta T_3^{\rm SVD}(n) + 
\delta T_3^\bot(n)$.

The order-by-order expansions of the cluster amplitudes and multipliers are found by inserting Eqs. 
(\ref{formalt})--(\ref{formall}) into Eqs. (\ref{delta12})--(\ref{mult3bot}). Subsequently, all 
terms of the same total order are grouped together and equated to zero. In this way one immediately 
finds that there are no zeroth-order 
contributions to the cluster amplitudes and multipliers, i.e. $\delta T(0) = 0$ and $\delta 
\mathcal{L}(0) = 0$. Moreover, by analyzing Eqs. (\ref{delta12})--(\ref{delta3bot}) one concludes 
that
the only first-order contribution to the perturbed cluster amplitudes is $\delta T_3^\bot(1)$ 
obtained from
\begin{align}
\label{t3_1}
 \langle\mu_3^\bot | \Big[ F, \delta T_3^\bot(1) \Big] \rangle + \langle \mu_3^\bot | \overline{W} 
\rangle = 0.
\end{align}
The remaining first-order components vanish, i.e. $\delta T_1(1)=\delta T_2(1)=\delta T_3^{\rm 
SVD}(1)=0$.
This means that the ``relaxation'' of the amplitudes corresponding to excitations already included 
in the SVD-CCSDT theory is of secondary importance and does not enter in the first order in the 
fluctuation potential. Similar conclusions hold for the multipliers where the only first-order 
contribution is $\delta \mathcal{L}_3^\bot(1)$ and reads
\begin{align}
 \langle\delta \mathcal{L}_3^\bot(1)|\Big[ F, \mu_3^\bot \Big] \rangle +
 \langle\mathcal{L}_{\rm SVD}|\Big[ \overline{W}, \mu_3^\bot \Big] \rangle = 0,
\end{align}
and $\delta\mathcal{L}_1(1)=\delta\mathcal{L}_2(1)=\delta \mathcal{L}_3^{\rm SVD}(1)=0$. In the 
second order we obtain the following expressions for the perturbed cluster amplitudes
\begin{subequations}
 \begin{align}
\begin{split}
 \langle\mu_m | \Big[ F, \delta T_m(2) \Big] \rangle +
 \langle\mu_m|\Big[ \overline{W}, \delta T_3^\bot(1) \Big] \rangle = 0,
 \;\;\; m=1,2,
\end{split}\\
\begin{split}
 \langle\mu_3^{\rm SVD} | \Big[ F, \delta T_3^{\rm SVD}(2) \Big] \rangle +
 \langle\mu_3^{\rm SVD}|\Big[ \overline{W}, \delta T_3^\bot(1) \Big] \rangle = 0,
\end{split}\\
\begin{split}
 \langle\mu_3^\bot | \Big[ F, \delta T_3^\bot(2) \Big] \rangle +
 \langle\mu_3^\bot|\Big[ \overline{W}, \delta T_3^\bot(1) \Big] \rangle = 0,
\end{split}
\end{align}
\end{subequations}
where we see, for the first time, a non-vanishing ``relaxation'' contribution. The second-order 
contributions to the multipliers are similarly found from the equations
\begin{align}
 \langle\delta \mathcal{L}_m(2)|\Big[ F, \mu_m \Big] \rangle +
 \langle\delta \mathcal{L}_3^\bot(1)|\Big[ \overline{W}, \mu_m \Big] \rangle + 
 \langle\mathcal{L}_{\rm SVD}|
 \Big[\big[ \overline{W},\delta T_3^\bot(1) \big], \mu_m\Big]\rangle = 0,
 \;\;\; m=1,2,
\end{align}
and analogously for the $\delta \mathcal{L}_3^{\rm SVD}(2)$ component. The second-order 
contribution 
to $\delta \mathcal{L}_3^\bot$ vanishes, i.e. $\delta \mathcal{L}_3^\bot(2)=0$, because the only 
relevant second-order term in Eq. (\ref{mult3bot}), namely $\langle\mathcal{L}_{\rm SVD}|
 \Big[\big[ \overline{W},\delta T_3^\bot(1) \big], \mu_3^\bot\Big]\rangle$, is zero due to 
conflicting excitation levels in bra and ket. 

The major practical advantage of the order-by-order 
expansion in comparison with the initial formulation given by Eqs. 
(\ref{delta12})--(\ref{mult3bot}) is the fact that at each level the highest-order contribution to 
$\delta \mathcal{L}$ and $\delta T$ appears only in the commutator with the Fock operator, e.g. 
$\langle\mu_3^\bot | \Big[ F, \delta T_3^\bot(n) \Big] \rangle$ or $\langle\delta 
\mathcal{L}_3^\bot(n)|\Big[ F, \mu_3^\bot \Big] \rangle$. This allows to invert each equation 
explicitly in a one step procedure, in contrast to Eqs. (\ref{delta12})--(\ref{mult3bot}) which 
require an iterative procedure to solve.

Having determined the order-by-order expansion of $\delta T$ and $\delta \mathcal{L}$ we 
proceed to the derivation of the corresponding energy corrections based on Eq. (\ref{lag1}). Since 
there are no zeroth-order contributions in $\delta T$ and $\delta \mathcal{L}$ one can easily show 
that $\delta E(0)=0$. Moreover, an inspection of Eq. (\ref{lag1}) proves that $\delta E(1)$ also 
vanishes and hence the energy corrections start at the second order in the 
fluctuation potential. To derive the higher-order corrections one has to keep 
in mind that the expression for $\delta E$ constructed above is variational with respect to $\delta 
T$ and $\delta \mathcal{L}$. As a result, the corrections $\delta E(n)$ obey the so-called Wigner 
rules. These rules state that for a given $n$, $\delta T(n)$ is sufficient to calculate corrections up to
$E(2n+1)$, while $\delta \mathcal{L}(n)$ -- up to $\delta E(2n+2)$. Guided by these rules we 
find the following expressions for $\delta E(n)$ up to the fourth order:
\begin{subequations}
 \begin{align}
\begin{split}
 \label{de2}
 \delta E(2) = 
 \langle \mathcal{L}_{\rm SVD} | \Big[ \overline{W}, \delta T_3^\bot(1) \Big] \rangle,
\end{split}\\
\begin{split}
 \label{de3}
 \delta E(3) = 
 \langle \mathcal{L}_3^\bot(1) | \Big[ \overline{W}, \delta T_3^\bot(1) \Big] \rangle,
\end{split}\\
\begin{split}
\label{de4}
 \delta E(4) = 
 \langle \mathcal{L}_3^\bot(1) | \Big[ \overline{W}, \delta T(2) \Big] \rangle +
 \langle \mathcal{L}_{\rm SVD} | \Big[ \big[ \overline{W}, \delta T_3^\bot(1) \big],
 \delta T_1(2) + \delta T_2(2) \Big]\rangle.
\end{split}
\end{align}
\end{subequations}
Note that the aforementioned ``relaxation'' of the cluster amplitudes, represented by $\delta 
T_1(2)$, $\delta T_2(2)$, and $\delta T_3^{\rm SVD}(2)$ operators, contributes for the first time 
in surprisingly high orders. The components $\delta T_1(2)$ and $\delta T_2(2)$ appear for the 
first time in $\delta E(4)$, while $\delta T_3^{\rm SVD}(2)$ does not enter until the fifth order.

In this work we concentrate on the leading-order correction to the SVD-CCSDT energy, $\delta E(2)$, 
given by Eq. (\ref{de2}). In this paragraph we bring this equation to an explicit form, more suitable
for further manipulations. First, note that Eq. (\ref{de2}) can be equivalently rewritten as
\begin{align}
\label{e2_2}
  \delta E(2) = 
 \langle \mathcal{L}_{\rm SVD} | \Big[ \overline{H}, \delta T_3^\bot(1) \Big] \rangle,
\end{align}
because $\overline{H}=\overline{F}+\overline{W}$, and $\langle \mathcal{L}_{\rm SVD} | \Big[ \overline{F}, \delta 
T_3^\bot(1) \Big] \rangle=\langle \mathcal{L}_{\rm SVD} | \Big[ F, \delta 
T_3^\bot(1) \Big] \rangle=0$. The latter equality is valid due to the condition (\ref{prop3}) and conflicting 
excitation levels in bra and ket. Next, we observe that Eq. (\ref{t3_1}) which defines the operator $\delta 
T_3^\bot(1)$ can be simplified to the form
\begin{align}
\label{t3_1_1}
 \langle\mu_3 | \Big[ F, \delta T_3(1) \Big] \rangle + \langle \mu_3 | \overline{H} \rangle = 0.
\end{align}
To derive this equation we exploited the facts that $\langle\mu_3^\bot | 
\overline{W}\rangle=\langle\mu_3^\bot | \overline{H}\rangle$ and $\langle\mu_3^\bot | 
\overline{H}\rangle = \langle\mu_3 | \overline{H}\rangle$. The former relationship is a direct 
consequence of Eq. (\ref{prop3}) while the latter holds due to the SVD-CCSDT stationarity 
condition, i.e. $\langle\mu_3^{\rm SVD}|\overline{H}\rangle=0$. The advantage of Eq. (\ref{t3_1_1}) in comparison
to Eq. (\ref{t3_1}) is that it can be explicitly solved since the Fock operator is diagonal in
the basis of canonical molecular orbitals. By combining Eqs. (\ref{e2_2}) and (\ref{t3_1_1}) we arrive at
\begin{align}
\label{fin1}
 \delta E(2) = -\left(\epsilon_{ijk}^{abc}\right)^{-1}\,
 \langle \mathcal{L}_1+\mathcal{L}_2+\mathcal{L}_3^{\rm SVD}|\overline{H}|_{ijk}^{abc}\rangle
 \langle_{ijk}^{abc}|\overline{H}\rangle.
\end{align}
Note that the manipulations outlined above allowed to remove the complementary subspace $\mu_3^\bot$ from the final 
working expression, eliminating the need to explicitly find the basis of $\mu_3^\bot$. However, this may no longer be 
possible in higher orders.

The equation (\ref{fin1}) constitutes the backbone of our formalism. However, for pragmatic 
reasons we introduce two additional approximations. They are not necessary to 
make the method practically feasible, but nonetheless they reduce the cost of the calculations 
considerably without sacrificing much accuracy.
First, we replace the $\mathcal{L}$ amplitudes by the corresponding $T_{\rm SVD}$ cluster 
amplitudes, i.e., we set $\mathcal{L}_n=T_n$, $n=1,2$, and $\mathcal{L}_3^{\rm SVD}=T_3^{\rm SVD}$, 
as is the usual practice in deriving non-iterative corrections accounting for higher-order 
excitations. This approximation eliminates the need to compute the SVD-CCSDT Lagrangian 
multipliers which is comparably expensive to the SVD-CCSDT calculation itself. Nonetheless, we 
note that the inclusion of the Langrangian amplitudes in other non-iterative methods, such as 
$\Lambda$-CCSD(T)~\cite{crawford98,kucharski98b,gwaltney01},
has been studied. An implementation of a related formalism in the SVD-CCSDT 
context is an interesting topic for a future work.

The second approximation is the neglect of the $\mathcal{L}_3^{\rm SVD}$ component in 
Eq. (\ref{fin1}). We found that this term is numerically negligible in most cases, especially for 
smaller SVD subspaces. At the same time the calculation of this term, while possible to accomplish 
with a $N^7$ scaling, 
is technically complicated and possesses a rather large prefactor. Further in the 
paper we demonstrate that the omission of this term results in a method that is already capable of 
reaching sufficient 
accuracy levels. After taking into account the aforementioned approximations we arrive at the final 
formula
\begin{align}
\label{fin2}
 \delta E_{\rm T+} = -\left(\epsilon_{ijk}^{abc}\right)^{-1}\,
 \langle T_1+T_2|\overline{H}|_{ijk}^{abc}\rangle
 \langle_{ijk}^{abc}|\overline{H}\rangle.
\end{align}
For the brevity, the method that adds the correction (\ref{fin2}) on top of the converged SVD-CCSDT 
energy is called SVD-CCSDT+ further in the text.

It is important to discuss two extreme cases of the formula (\ref{fin2}): the case when the 
SVD subspace is empty, $\mu_3^{\rm SVD}=\varnothing$, and the case when it spans the whole space of 
triple excitations, $\mathrm{span}(\mu_3)=\mathrm{span}(\mu_3^{\rm SVD})$. The former case 
trivially corresponds to the CCSD 
calculations where $T_3^{\rm SVD}=0$ and the expression for the correction simplifies to
\begin{align}
 \delta E_{\rm T+} = -\left(\epsilon_{ijk}^{abc}\right)^{-1}\,
 \langle T_1+T_2|\widetilde{F}+\widetilde{W}|_{ijk}^{abc}\rangle
\langle_{ijk}^{abc}|\left[\widetilde{W},T_2\right]+\half\left[\left[\widetilde{F}+\widetilde{W},
T_2\right],T_2\right] 
\rangle,
\end{align}
where $\widetilde{F}=e^{-T_1} F e^{T_1}$. If we additionally dropped all terms that are of 
quadratic and higher order in the cluster amplitudes we would obtain the expression defining the 
CCSD(T) theory. This means that for an empty SVD subspace results obtained with the SVD-CCSDT+ 
method should be close to the CCSD(T) theory, assuming that we are dealing with well-behaved 
systems where the norm of the cluster amplitudes is significantly smaller than the unity. In the 
second extreme case ($\mu_3=\mu_3^{\rm SVD}$) it is straightforward to show that the $\delta E_{\rm 
T+}$ correction is rigorously equal to 
zero. This is a consequence of the fact that the term $\langle_{ijk}^{abc}|\overline{H}\rangle$ in 
Eq. (\ref{fin2}), defining the CCSDT stationary condition, vanishes for $\mu_3=\mu_3^{\rm SVD}$. 
This property of the $\delta 
E_{\rm T+}$ correction shows that the formula (\ref{fin2}) does not introduce any spurious 
double-counting of excitations which would degrade the 
accuracy. However, 
note that if the term $\langle_{ijk}^{abc}|\overline{H}\rangle$ in Eq. (\ref{fin2}) was 
approximated in 
any way, the zero limit of the correction would no longer be strictly guaranteed.

Finally, we discuss the computational cost of evaluating Eq. (\ref{fin2}) and some details of the 
implementation. Explicit expressions for the residual 
$\langle_{ijk}^{abc}|\overline{H}\rangle$ expressed through cluster amplitudes and molecular 
integrals were given in Ref.~\onlinecite{noga87} and there is no point in repeating them here. 
Therefore, for 
illustrative purposes we 
concentrate only on a single term in $\langle_{ijk}^{abc}|\overline{H}\rangle$ that determines the 
overall scaling of the method. It reads
\begin{align}
\label{highterm}
 \left( 1 + P_{bj,ck} \right) \left( 1 + P_{ai,bj} + P_{ai,ck}\right) 
\big[\chi_{bd}^{ce}\,t_{ijk}^{ade}\big],
\end{align}
where $\chi_{bd}^{ce}$ is an intermediate quantity given by Eq. (13) in Ref.~\onlinecite{lesiuk20} 
and
the symbol $P_{ai,bj}$ denotes the permutation operator that exchanges pairs of 
indices $i\leftrightarrow j$ and $a\leftrightarrow b$ simultaneously. Without any simplifications 
the 
computational cost of this term scales as $O^3V^5\propto N^8$. However, by exploiting the 
decomposition format of the triply-excited amplitudes, Eq. (\ref{tuck1}), and properly arranging 
the order of tensor contractions the assembly of Eq. (\ref{highterm}) can be decomposed into a 
series of steps
\begin{align}
 \left( 1 + P_{bj,ck} \right) \left( 1 + P_{ai,bj} + P_{ai,ck}\right)
 \bigg[U_{ai}^X \Big[\big[\big( \chi_{bd}^{ce}\,U_{ek}^Z \big)\,U_{dj}^Y\big] t_{XYZ}\Big]\bigg],
\end{align}
where the parentheses indicate the sequence of operations. By recalling that 
$N_{\mathrm{SVD}}$ scales linearly with the system size one can show that each step scales as $N^7$ 
or less. The most expensive is the third step (counting from the innermost parentheses) scaling as 
$O^2 N_{\mathrm{SVD}}^3 V^2$. In order to avoid memory bottlenecks the 
quantity $\langle_{ijk}^{abc}|\overline{H}\rangle$ is evaluated on-the-fly in batches with three 
fixed occupied indices ($ijk$). Each batch is immediately consumed in evaluation of the 
respective contribution to $\delta E_{\rm T+}$, Eq. (\ref{fin2}), and then discarded. 
This part of the algorithm is similar to the conventional scheme employed for evaluation of the (T) 
correction (see, for example, Refs.~\onlinecite{lee90,rendell91,jankowski08}) and possesses the 
same 
computational cost, namely $O^3V^4\propto N^7$. For small SVD subspace size, this cost is dominant 
in 
comparison with $O^2 N_{\mathrm{SVD}}^3 V^2$ necessary to assemble 
$\langle_{ijk}^{abc}|\overline{H}\rangle$. However, in more accurate calculations one expects that 
$N_{\mathrm{SVD}}\approx V$ and the computation of $\langle_{ijk}^{abc}|\overline{H}\rangle$ 
becomes the limiting step. In this regime it is also worthwhile to compare the cost of evaluating 
the SVD-CCSDT+ correction with a single SVD-CCSDT iteration. The latter is characterized by the 
$OV^2N_{\mathrm{SVD}}^3$ scaling, as discussed in Ref.~\onlinecite{lesiuk20}, by a factor of $O$ 
smaller than the computation of $\langle_{ijk}^{abc}|\overline{H}\rangle$. Therefore, 
the total cost of computing the $\delta E_{\rm T+}$ correction is predicted to be roughly 
comparable to $O$ SVD-CCSDT iterations, assuming that the prefactors are of a similar magnitude and 
that $N_{\mathrm{SVD}}\approx V$. A detailed comparison of timings obtained for realistic systems 
is given further in the text. In Supporting Information we investigate the computational complexity of evaluating the 
$\delta E_{\rm T+}$ correction for a model system: linear alkanes with increasing chain length. Direct comparison of
computational timings obtained reveals a slightly lower scaling ($N^{6.37}$) than predicted theoretically ($N^7$). 
This deviation is due to terms in the $\delta E_{\rm T+}$ correction that can be evaluated with $N^5$ or $N^6$ cost, 
but have a relatively large prefactor. Nonetheless, as the system size is increased further, the cost of such terms is 
going to decrease (on a relative basis), leading to the $N^7$ scaling of the method.

\section{Numerical results and discussion}
\label{sec:numer}

\subsection{Computational details}
\label{subsec:kompot}

Unless explicitly stated otherwise, all calculations reported in this work employ the Dunning-type 
cc-pVTZ basis set~\cite{dunning89}. The corresponding density-fitting auxiliary basis set 
cc-pVTZ-MP2FIT was taken from the work of Weigend et al~\cite{weigend98,weigend02}. Pure spherical 
representation ($5d$, $7f$, etc.) of both basis sets was employed. All theoretical methods 
described in this work were implemented in a locally modified version of the \textsc{Gamess} 
program package~\cite{gamess1}.

Density-fitting was employed by default at every stage of calculations apart from solving the 
Hartree-Fock equations where the exact two-electron integrals were used. Additionally, 
reference CCSDT results were obtained using the exact integrals because, to the best of our 
knowledge, no DF-CCSDT implementation is currently available. The uncompressed CCSDT calculations 
were performed with the help of the \textsc{CFour} program package~\cite{cfour1,cfour2}. In all 
correlated calculations we employ the frozen-core approximation by dropping $1s^2$ core orbitals of 
the first-row atoms (Li--Ne).

\subsection{Numerical verification of the condition (\ref{prop3})}
\label{subsec:gbctest}

\begin{figure}[t]
 \includegraphics[scale=0.75]{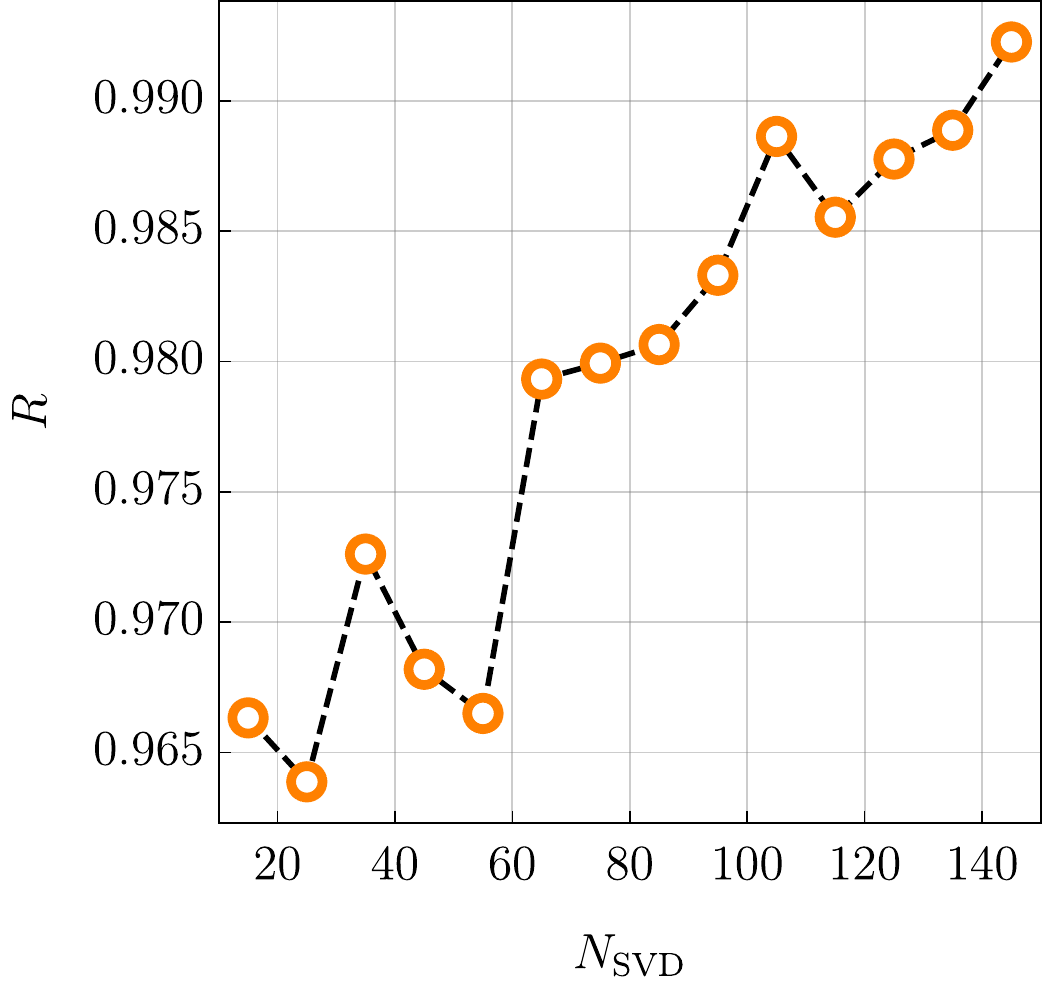}
 \caption{\label{gbc} Values of the $R$ coefficient, see Eq. (\ref{rcoef}), as a function of the 
 SVD subspace size for the HF molecule (internuclear distance $0.917\,$\AA{}, cc-pVTZ basis 
 set). The black dashed lines are linear functions connecting two neighboring data points.}
\end{figure}

The derivation of the SVD-CCSDT+ correction presented in Sec. \ref{subsec:pert} relies on the 
assumption that the SVD subspace is 
closed under the action of the Fock operator. When the SVD subspace is incomplete this constitutes 
an approximation whose quality has to be verified numerically. To quantify the accuracy of Eq. 
(\ref{gbc1}) 
we consider the action of the Fock operator on the sum of all SVD vectors, i.e. $\sum_{XYZ} 
F|XYZ\rangle$. For brevity, we introduce the symbol $\Psi_{\rm SVD} = \sum_{XYZ} |XYZ\rangle$.
The function $F\Psi_{\rm SVD}$ is projected separately onto the SVD subspace and onto the full 
space of triple excitations. Finally, square norms of both projections are formed and the square 
root of their ratio is calculated. The resulting quantity measures the magnitude of the component 
of 
$F\Psi_{\rm SVD}$ that resides within the SVD subspace in relation to the total norm of the 
$F\Psi_{\rm SVD}$ function. Expressed mathematically, this reads
\begin{align}
\label{rcoef}
 R = \sqrt{\frac{\langle \Psi_{\rm SVD}|F|XYZ\rangle\langle XYZ|F|\Psi_{\rm 
SVD}\rangle}{\langle \Psi_{\rm SVD}|F|_{ijk}^{abc}\rangle\langle_{ijk}^{abc}|F|\Psi_{\rm 
SVD}\rangle}}.
\end{align}
By construction, the coefficient $R$ takes values between $0$ and $1$. In a situation when Eq. 
(\ref{prop3}) is satisfied exactly one strictly has $R=1$. Therefore, the deviation of the 
coefficient $R$ from the unity is a quantitative measure of the accuracy of the 
condition~(\ref{prop3}). However, it is important to point out that is not guaranteed that the 
coefficient $R$ vanishes monotonically as the SVD subspace size is increased.

As an illustrative example, we calculated the coefficient $R$ as a function of the SVD subspace 
size for the hydrogen fluoride (HF) molecule (internuclear distance $0.917\,$\AA{}). 
For this system the maximum size of the SVD subspace equals to $195$. The results are presented in 
Fig. \ref{gbc}. The first important observation is that even for small SVD subspaces 
($N_{\mathrm{SVD}} 
\approx 10$) the values of the coefficient $R$ are already larger than $0.95$. This 
further increases above $0.98$ when a larger number of SVD vectors are included 
($N_{\mathrm{SVD}} \approx 50$). Therefore, for SVD subspaces large enough to 
be practically useful, the error resulting from Eq. (\ref{gbc1}) should not exceed 
a few percent. Considering other possible sources of error, this is acceptable from the point of 
view of the present work. In Supporting Information we present additional numerical results 
analogous to Fig. \ref{gbc} for other molecules. The conclusions of these calculations are 
essentially the same as discussed above.

\subsection{Accuracy of the SVD-CCSDT+ method: \\ total correlation energies}
\label{subsec:plustot}

Before we present calculations of chemically-relevant quantities for larger molecular systems, it is advantageous to 
study errors of the SVD-CCSDT and SVD-CCSDT+ methods in reproduction of total correlation energies taking the exact 
CCSDT method as the reference. To this end we selected a set of $16$ small molecules comprising 
$2-5$ atoms. The list 
of the molecules together with their geometries in the Cartesian format are given in the Supporting Information. For 
each molecule we performed SVD-CCSDT and SVD-CCSDT+ calculations (cc-pVTZ basis set) with the size 
of the SVD subspace being linearly 
related to the total number of orbitals ($N_{\mathrm{MO}}$) in a given system, that is $N_{\mathrm{SVD}}=x\cdot 
N_{\mathrm{MO}}$. We consider several representative values of the $x$ parameter, namely 
$x=\frac{1}{6}$, $\frac{1}{3}$, $\frac{1}{2}$, $\frac{2}{3}$, $\frac{5}{6}$, $1$. To 
minimize the impact of the density-fitting approximation on the computed correlation energies, we employ a large 
cc-pV5Z-RI auxiliary basis set. With this setup the results are virtually free of the density-fitting error which does 
not exceed a few parts per million in all cases.

\begin{figure}[ht]
\includegraphics[scale=1.00]{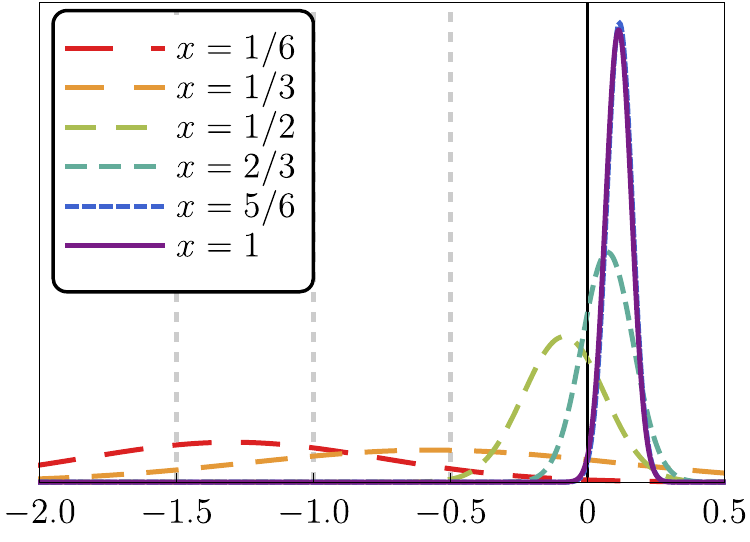}
\includegraphics[scale=1.00]{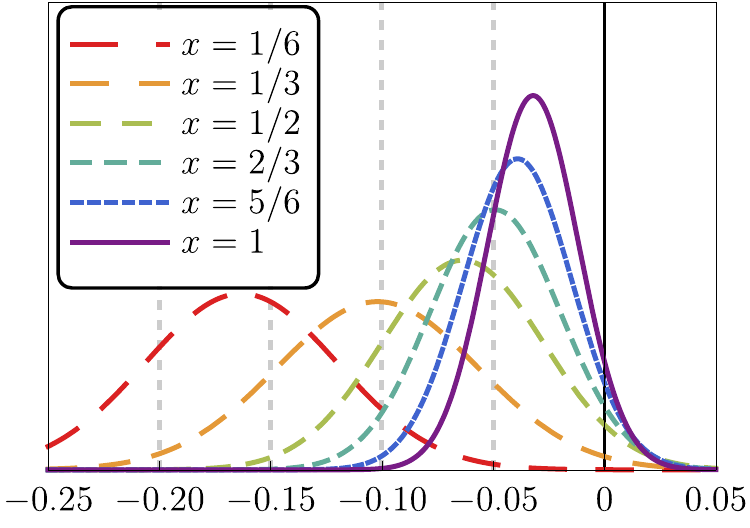}
\caption{\label{fig:total} 
Distribution of relative errors (in percent) in the SVD-CCSDT (left panel) and SVD-CCSDT+ (right 
panel) correlation energies with respect to the exact CCSDT method (cc-pVTZ basis set). The 
parameter $x$ defines the size of the triple excitation subspace, $N_{\mathrm{SVD}}=x\cdot 
N_{\mathrm{MO}}$, where $N_{\mathrm{MO}}$ is the total number of orbitals in a given system. Notice 
the change of scale on the horizontal axis.
}
\end{figure}

Further in this section we adopt the \emph{relative} error in the correlation energy, defined as
\begin{align}
\label{errort}
 \Delta_i = \frac{E_{\mathrm{method},i}-E_{\mathrm{CCSDT},i}}{E_{\mathrm{CCSDT},i}},
\end{align}
as the size-intensive measure of the quality of the results. In Eq. (\ref{deltat}) the index $i$ enumerates the 
molecules in the test set and the symbol ``method'' refers to either SVD-CCSDT or SVD-CCSDT+. To 
perform a statistical analysis of the results we calculated the mean relative error and its 
standard deviation
\begin{align}
 &\bar{\Delta}=\frac{1}{n}\sum_{i=1}^n \Delta_i, \\
 &\Delta_{\mathrm{std}}^2=\frac{1}{n-1}\sum_{i=1}^n \left( \Delta_i -\bar{\Delta} \right)^2,
\end{align}
where $n=16$ is the number of molecules in the test set. We found that the statistical distribution 
of the relative 
error is approximately normal for each value of the $x$ parameter. Therefore, to simplify the presentation the 
quantities $\bar{\Delta}$ and $\Delta_{\mathrm{std}}$ are represented graphically in Fig. 
\ref{fig:total} in terms of the Gaussian functions
\begin{align}
 \rho(\Delta) = \mathcal{N}e^{-\left( \Delta -\bar{\Delta} \right)^2/2\Delta_{\mathrm{std}}^2},
\end{align}
where $\mathcal{N}$ is a constant chosen such that $\rho(\Delta)$ is normalized to the unity. Raw 
values of the quantities $\bar{\Delta}$ and $\Delta_{\mathrm{std}}$ for all $x$ under consideration 
are given in the Supporting Information.

The results represented in Fig. \ref{fig:total} show that the SVD-CCSDT method has a tendency to 
underestimate the correlation energies for small $x$, but then it overshoots the exact results as 
$x$ is increased. For $x=1$ the mean error of the SVD-CCSDT method is slightly above 0.1\% and, as 
demonstrated in Ref.~\onlinecite{lesiuk20}, further increase of the parameter $x$ leads to a 
smooth, albeit slower, convergence towards the exact value. The behavior of the SVD-CCSDT+ is both 
qualitatively and quantitatively different. First, the overshooting tendency is absent and the 
convergence is smooth starting with the smallest $x$ considered here. Second, the magnitude of the 
error is reduced considerably; for example, for $x=1$ the SVD-CCSDT+ method achieves the mean 
relative error of about $-$0.03\% which is by a factor of about four smaller than SVD-CCSDT. It 
is worth point out that for all molecules considered here, the SVD-CCSDT+ results approach their 
limit from above in a regular fashion. This opens up a possibility of extrapolation similar as in 
Refs.~\onlinecite{landau10,marti18}, but a detailed analysis of this problem is beyond the scope of 
the present work.

\subsection{Accuracy of the SVD-CCSDT+ method: relative energies}
\label{subsec:plustest}

\begin{figure}[t]
 \includegraphics[scale=0.65]{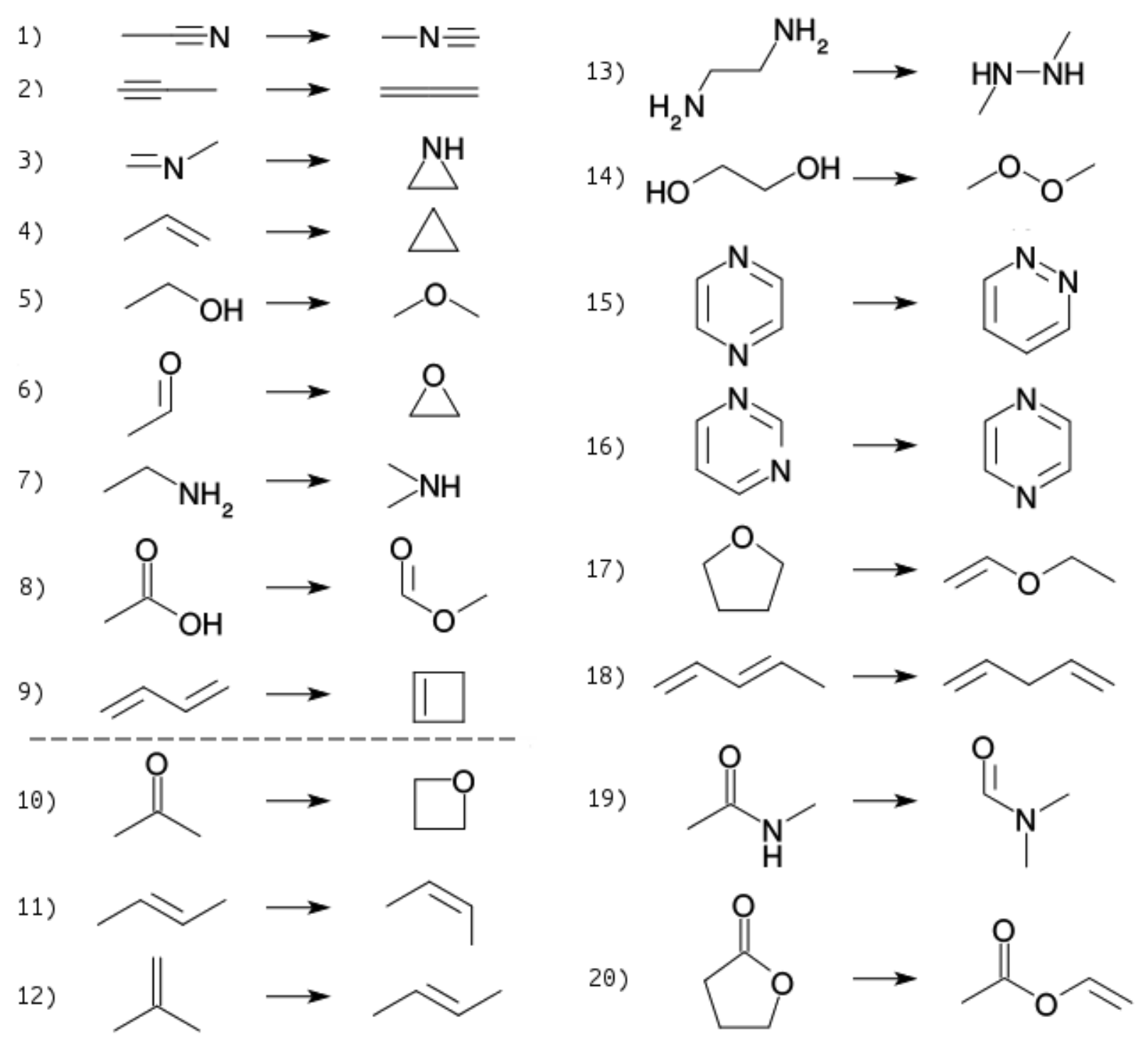}
 \caption{\label{fig:react} List of isomerization reactions considered in the present work. The 
horizontal dashed line separates two groups of reactions, see the discussion in Sec. 
\ref{subsec:plustest}.}
\end{figure}

\begin{table}[!ht]
  \caption{Isomerization energies (in kJ/mol) for systems presented in Fig.~\ref{fig:react} 
calculated with the SVD-CCSDT and SVD-CCSDT+ methods as a function of the SVD subspace size 
($N_{\mathrm{SVD}}$). For each reaction, the SVD-CCSDT results are given in the first row while the 
SVD-CCSDT+ results in the second row. The corresponding results obtained with the exact CCSDT method 
are given in the last column. The number of (active) occupied/virtual orbitals in each system is given in 
parentheses below the reaction numbers.}
  \label{tab:isomers}
  \begin{tabular}{clrrrrrrrr}
    \hline\hline
    no. &  & \multicolumn{7}{c}{$N_{\mathrm{SVD}}$} & CCSDT \\
    \hline
                      & \multicolumn{1}{c}{$20$}
                      & \multicolumn{1}{c}{$40$}
                      & \multicolumn{1}{c}{$60$}
                      & \multicolumn{1}{c}{$80$}
                      & \multicolumn{1}{c}{$100$} 
                      & \multicolumn{1}{c}{$120$}
                      & \multicolumn{1}{c}{$140$}
                      & \multicolumn{1}{c}{$160$} \\
    \hline
    1       & 105.50 & 102.21 & 101.11 & 101.47 & 101.02
            & 101.05 & 101.10 & 100.95 & \multirow{2}{*}{100.79} \\
    (8/121) & 100.70 & 100.46 & 100.66 & 100.69 & 100.78
            & 100.75 & 100.72 & 100.72 \\
    \hline
    2       & 8.69 & 7.03 & 6.99 & 6.49 & 5.33 
            & 5.55 & 5.37 & 5.18 & \multirow{2}{*}{4.43} \\
    (8/135) & 4.55 & 4.31 & 4.34 & 4.47 & 4.58
            & 4.53 & 4.52 & 4.52 \\
    \hline
    3       & 49.65 & 46.03 & 46.61 & 46.57 & 45.90
            & 46.06 & 45.88 & 45.92 & \multirow{2}{*}{45.99} \\
    (9/148) & 45.95 & 45.80 & 45.78 & 45.81 & 45.86
            & 45.88 & 45.88 & 45.89 \\
    \hline
    4       & 35.31 & 32.72 & 33.64 & 32.85 & 32.29
            & 32.47 & 32.40 & 32.30 & \multirow{2}{*}{32.13} \\
    (9/162) & 32.20 & 32.00 & 31.96 & 31.95 & 31.98
            & 31.98 & 32.04 & 32.05 \\
    \hline
    5        & 49.86 & 49.76 & 50.00 & 50.00 & 49.66
             & 49.19 & 49.00 & 49.00 & \multirow{2}{*}{48.80} \\
    (10/161) & 48.88 & 48.87 & 48.85 & 48.84 & 48.79
             & 48.77 & 48.78 & 48.77 \\
    \hline
    6       & \multicolumn{1}{c}{---\textsuperscript{a}}
            & 114.03 & 114.16 & 113.67 & 113.12
            & 112.90 & 112.69 & 112.62 & \multirow{2}{*}{112.44} \\
    (9/134) & \multicolumn{1}{c}{---\textsuperscript{a}}
            & 112.48 & 112.45 & 112.38 & 112.40
            & 112.38 & 112.38 & 112.37 \\
    \hline
    7        & 34.15 & 34.52 & 34.41 & 34.27 & 31.15
             & 33.98 & 33.86 & 33.89 & \multirow{2}{*}{33.67} \\
    (10/175) & 33.74 & 33.73 & 33.76 & 33.68 & 33.68
             & 33.66 & 33.66 & 33.65 \\
    \hline
    8        & 69.77 & 70.79 & 70.20 & 70.33 & 69.73
             & 69.68 & 69.44 & 69.36 & \multirow{2}{*}{69.12} \\
    (12/160) & 69.29 & 69.24 & 69.22 & 69.19 & 69.17
             & 69.16 & 69.15 & 69.14 \\
    \hline
    9        & 52.44 & 50.32 & 49.67 & 50.52 & 49.63
             & 48.88 & 48.80 & 48.82 & \multirow{2}{*}{48.85} \\
    (11/189) & 48.93 & 48.85 & 48.80 & 48.70 & 48.65
             & 48.60 & 48.67 & 48.69 \\
    \hline\hline
  \end{tabular}
  \vspace{-0.25cm}
\begin{flushleft}
\small \textsuperscript{a} SVD-CCSDT iterations did not converge.
\end{flushleft}
\end{table}

In the previous we have considered the accuracy of the SVD-CCSDT+ method in reproduction of the 
total correlation energies. However, relative quantities are of prime importance in most 
practical situations. To analyze the accuracy and computational efficiency of the proposed method 
in the latter case we calculated isomerization 
energies for $20$ organic systems from the set previously considered by Grimme et 
al~\cite{grimme07}. This set covers a wide 
range of structural changes -- from relatively simple cis-trans isomerizations to migrations
of whole functional groups. Besides several reactions involving only hydrocarbons, we include 
examples 
that also involve heteroatoms (oxygen and nitrogen) to cover the most common bonding situations 
found in
organic molecules. The range of isomerization energies is also broad, spanning from only a few 
kJ/mol to 
hundreds of kJ/mol. The list of isomerization reactions is given in Fig.~\ref{fig:react}. They are 
arranged roughly in the order of increasing system size. The largest molecule considered in this 
work (reaction 20) contains 46 electrons (34 correlated) with 264 functions in the atomic basis set 
and 666 functions in the density-fitting basis set. All reactions given in Fig.~\ref{fig:react} are 
formulated such that the isomerization energies are positive.

The isomerization reactions given in Fig.~\ref{fig:react} are divided into two groups. The first 
group (reactions 1--9) involve relatively small molecular systems for which full (uncompressed) 
CCSDT computations are available. This allows to ascertain the accuracy of the 
SVD-CCSDT+ method unambiguously. The second group (reactions 11--20) includes also larger species for which
the exact CCSDT calculations are not practical with our computational resources. For each isomerization reaction from 
Fig.~\ref{fig:react} we carried out 
SVD-CCSDT calculations followed by evaluation of the SVD-CCSDT+ correction. The size of the SVD 
subspace was varied from $N_{\mathrm{SVD}}=20$ up to $N_{\mathrm{SVD}}=160$ in steps of $20$. For 
the second group of reactions we additionally obtained results for $N_{\mathrm{SVD}}=180$ and 
$N_{\mathrm{SVD}}=200$ to fully converge the energy differences. The results for the first and 
second group of reactions are given in Tables~\ref{tab:isomers} and {\ref{tab:isomers2}, 
respectively.

We begin our analysis by considering the first group of reactions, see Table~\ref{tab:isomers}. 
Similarly as in the previous work, the SVD-CCSDT results converge rather quickly to 
the accuracy level of a fraction of 1 kJ/mol with increasing SVD subspace size. Beyond this point 
the convergence slows down (reactions 5 and 6) or even becomes mildly oscillatory (reactions 4 
and 7). To reduce the error of the SVD-CCSDT method to the level of about 0.1 kJ/mol in 
all cases, a further increase of $N_{\mathrm{SVD}}$ would be needed. This is similar to the 
accuracy of the SVD-CCSDT method for the total correlation energies reported in Sec. 
\ref{subsec:plustot} and in Ref.~\onlinecite{lesiuk20}, albeit for the isomerization energies we do 
not observe the overshooting tendency.

By adding the SVD-CCSDT+ correction the results improve considerably. The accuracy of a fraction 
of 1 kJ/mol is achieved with significantly smaller SVD subspaces ($N_{\mathrm{SVD}}=40-60$). 
A further increase of $N_{\mathrm{SVD}}$ to $\approx N_{\mathrm{MO}}$ allows to stabilize the 
results to within 0.01--0.02 kJ/mol (or 0.02--0.03\% on the relative basis) in most cases. 
Therefore, the overall behavior of the SVD-CCSDT+ method in calculation of the isomerization 
energies is similar as in the case of the total energies reported in Sec. \ref{subsec:plustot}. 
However, the improvement over the SVD-CCSDT is larger which suggests that the SVD-CCSDT+ method 
benefits from a more systematic error cancellation.

A puzzling feature of the SVD-CCSDT+ results is that despite the apparently tight 
convergence to within 0.01--0.02 kJ/mol for $N_{\mathrm{SVD}}\approx N_{\mathrm{MO}}$, the true 
errors with respect to the exact CCSDT are larger. On a relative basis, this 
phenomenon is most pronounced for reaction 2 where the SVD-CCSDT+ results for 
$N_{\mathrm{SVD}}=160$ are converged to within 0.01 kJ/mol, but the deviation from the exact CCSDT 
is about 0.09 kJ/mol. Some of this difference can clearly be attributed to the perturbative nature 
of the SVD-CCSDT+ method and additional simplifications described in Sec.~\ref{subsec:pert}. 
However, the density-fitting approximation of the two-electron integrals may also be suspected to 
bring a significant contribution to the observed deviation. In fact, density-fitting errors of a 
similar magnitude were observed in the DF-CCSD(T) method, see Ref.~\onlinecite{deprince13} for a 
detailed analysis. To study this issue we 
repeated the SVD-CCSDT+ calculations for reaction 2 within the same orbital basis set, but 
increased the size of the auxiliary basis set by one cardinal number. The isomerization energy 
obtained in this way for $N_{\mathrm{SVD}}=160$ turned out to be 4.47 kJ/mol, compared with 4.52 
kJ/mol obtained previously. This reduced the deviation from the exact CCSDT result (4.43 kJ/mol) by 
a factor of two. Further increase of the size of the auxiliary basis leads to no appreciable 
improvement in the results. Therefore, we recommend to increase the size of the auxiliary basis set 
by one cardinal number in the SVD-CCSDT+ calculations if the accuracy of 0.1 kJ/mol or better is 
desired. Alternatively, other methods designed to reduce the density-fitting error can be 
adopted~\cite{schurkus17,lesiuk20b} or the density-fitting approximation can be entirely replaced 
by the Cholesky decomposition~\cite{beebe77,koch03,pedersen04,folkestad19} where a stricter error 
control is possible.

\begin{table}[!ht]
  \caption{Isomerization energies (in kJ/mol) for systems presented in Fig.~\ref{fig:react} 
calculated with the SVD-CCSDT and SVD-CCSDT+ methods as a function of the SVD subspace size 
($N_{\mathrm{SVD}}$). For each reaction, the SVD-CCSDT results are given in the first row while the 
SVD-CCSDT+ results in the second row. The number of (active) occupied/virtual orbitals in each system is given in 
parentheses below the reaction numbers.}
  \label{tab:isomers2}
  \begin{tabular}{crrrrrrrrrr}
    \hline\hline
    no. & \multicolumn{10}{c}{$N_{\mathrm{SVD}}$} \\
    \hline
                      & \multicolumn{1}{c}{$20$}
                      & \multicolumn{1}{c}{$40$}
                      & \multicolumn{1}{c}{$60$}
                      & \multicolumn{1}{c}{$80$}
                      & \multicolumn{1}{c}{$100$} 
                      & \multicolumn{1}{c}{$120$}
                      & \multicolumn{1}{c}{$140$}
                      & \multicolumn{1}{c}{$160$}
                      & \multicolumn{1}{c}{$180$}
                      & \multicolumn{1}{c}{$200$} \\
    \hline
    10       & 133.22 & 133.27 & 132.53 & 132.18 & 132.42
             & 131.54 & 131.46 & 131.23 & 131.27 & 131.02 \\
    (12/188) & 130.80 & 130.65 & 130.75 & 130.65 & 130.63
             & 130.54 & 130.53 & 130.53 & 130.54 & 130.53 \\
    \hline
    11       & 4.97 & 5.06 & 4.63 & 4.78 & 4.67 
             & 4.71 & 4.79 & 4.67 & 4.73 & 4.68 \\
    (12/216) & 4.57 & 4.69 & 4.66 & 4.61 & 4.65
             & 4.63 & 4.64 & 4.64 & 4.65 & 4.65 \\
    \hline
    12       & 3.90 & 4.45 & 4.90 & 4.19 & 4.66
             & 4.67 & 4.57 & 4.63 & 4.59 & 4.55 \\
    (12/216) & 4.59 & 4.66 & 4.66 & 4.67 & 4.67
             & 4.66 & 4.64 & 4.66 & 4.65 & 4.65 \\
    \hline
    13       & 118.61 & 117.90 & 117.66 & 117.68 & 117.63
             & 117.08 & 116.95 & 116.61 & 116.45 & 116.46 \\
    (13/215) & 116.44 & 116.28 & 116.25 & 116.20 & 116.17
             & 116.10 & 116.11 & 116.10 & 116.10 & 116.09 \\
    \hline
    14       & 273.71 & 270.06 & 271.01 & 270.74 & 269.29
             & 268.74 & 268.46 & 268.27 & 268.13 & 268.08 \\
    (13/187) & 268.00 & 267.62 & 267.62 & 267.45 & 267.42
             & 267.37 & 267.34 & 267.32 & 267.31 & 267.30 \\
    \hline
    15       & 77.34 & 76.29 & 76.00 & 76.46 & 75.56
             & 76.76 & 75.52 & 75.70 & 75.53 & 75.55 \\
    (15/215) & 75.66 & 75.57 & 75.59 & 75.55 & 75.57
             & 75.51 & 75.50 & 75.50 & 75.48 & 75.48 \\
    \hline
    16       & 19.89 & 18.37 & 20.51 & 19.74 & 20.88
             & 19.94 & 20.15 & 19.81 & 20.02 & 19.87 \\
    (15/215) & 18.99 & 19.37 & 19.41 & 19.49 & 19.44
             & 19.47 & 19.46 & 19.45 & 19.46 & 19.46 \\
    \hline
    17       & 49.24 & 49.70 & 50.08 & 49.95 & 50.63
             & 50.82 & 51.91 & 52.49 & 52.56 & 52.71 \\
    (15/242) & 52.70 & 52.84 & 52.90 & 52.98 & 53.02
             & 53.03 & 53.01 & 53.02 & 53.01 & 53.02 \\
    \hline
    18       & 27.75 & 24.90 & 25.20 & 25.54 & 25.60
             & 25.72 & 25.88 & 26.08 & 26.10 & 26.22 \\
    (14/243) & 26.98 & 26.79 & 26.78 & 26.85 & 26.82
             & 26.80 & 26.82 & 26.82 & 26.83 & 26.83 \\
    \hline
    19       & 39.52 & 40.35 & 40.14 & 39.54 & 39.69
             & 39.48 & 39.59 & 39.24 & 39.02 & 39.01 \\
    (15/228) & 38.60 & 38.70 & 38.61 & 38.63 & 38.60
             & 38.59 & 38.57 & 38.54 & 38.53 & 38.53 \\
    \hline
    20       & 58.98 & 59.41 & 59.21 & 59.03 & 59.45
             & 60.32 & 61.12 & 61.44 & 61.73 & 61.98 \\
    (17/241) & 62.43 & 62.42 & 62.56 & 62.67 & 62.73
             & 62.75 & 62.76 & 62.77 & 62.75 & 62.76 \\
    \hline\hline
  \end{tabular}
\end{table}

Among the isomerizations considered in Table~\ref{tab:isomers} the reaction 1 requires some 
additional attention. For this system an unexpectedly large discrepancy has been found between the 
CCSD(T) results published recently~\cite{grimme07} and the available experimental 
data~\cite{banghal77,xuwu83,pedley96,linstrom20}. While the CCSD(T) isomerization energy from 
Ref.~\onlinecite{grimme07} is 101.3 kJ/mol, the experimental result reads 89.1 kJ/mol. As argued in 
Ref.~\onlinecite{grimme07} this substantial difference of about 12 kJ/mol cannot be explained by 
effects such as the basis set incompleteness, core-valence correlation contributions, relativity or 
zero-point vibrational energy corrections. 
This led to the conclusion that the experimental uncertainty is the most likely source of the 
problem. However, it is worthwhile to analyze 
whether the post-CCSD(T) effects may explain the discrepancy. To this end we carried out additional 
CCSD(T) and SVD-CCSDT+ calculations using cc-pVQZ orbital basis set. We obtained $-0.77$ kJ/mol 
difference between CCSD(T) and SVD-CCSDT+ results for $N_{\mathrm{SVD}}=200$. This result is 
converged to within $0.01-0.02$ kJ/mol with respect to the size of the SVD subspace. Within the 
smaller cc-pVTZ basis set the analogous result reads $-0.56$ kJ/mol. 
To further minimize the finite basis set error we employed the two-point (cc-pVTZ/cc-pVQZ) 
Riemann extrapolation~\cite{lesiuk19b} towards the complete basis set limit. This gives the final 
estimation of $-0.97$ kJ/mol for the difference between CCSD(T) and CCSDT isomerization energies of 
reaction 1. Clearly, this value is too small to explain the observed difference between theory and 
experiment, supporting the conclusions of Ref.~\onlinecite{grimme07}. The example of reaction 1 
also demonstrates the usefulness of the SVD-CCSDT+ method in computation of the ``pure'' 
post-CCSD(T) effects which are known to be difficult to reproduce accurately for polyatomic 
molecules\cite{smith14}.

Finally, let us consider the second group of isomerization reactions. The results obtained with 
SVD-CCSDT and SVD-CCSDT+ methods are given in Table~\ref{tab:isomers2}. Since this group of 
reactions involves larger molecules, we additionally include results obtained for 
$N_{\mathrm{SVD}}=180$ and $N_{\mathrm{SVD}}=200$. In general, the picture drawn from the data 
presented in Table~\ref{tab:isomers2} is analogous as discussed above. In all cases, with 
$N_{\mathrm{SVD}}=160-200$ the SVD-CCSDT+ results are essentially converged to within 0.01--0.02 
kJ/mol. Taking other sources of error into account one can estimate that the difference with 
respect to the exact CCSDT method is, on average, smaller than 0.1 kJ/mol in this regime. At the 
same time, if the accuracy level of a few tenths of kJ/mol is sufficient, 
reliable results can be obtained with small SVD subspaces, $N_{\mathrm{SVD}}=20-40$. This is a 
considerable advantage of the SVD-CCSDT+ method over SVD-CCSDT as the latter is not trustworthy in 
the small $N_{\mathrm{SVD}}$ regime.

Last but not least, let us discuss the timings of the SVD-CCSDT+ calculations. As an 
illustrative 
example let us consider the product of reaction 15 (236 orbital basis set functions, 606 auxiliary 
basis set functions, 30 correlated electrons). For this system $N_{\mathrm{SVD}}=120$ is sufficient 
to converge the SVD-CCSDT+ results to within a few hundreds of kJ/mol. As a reference point, CCSD 
calculations for this system take approximately 520 min (converged within 18 iterations). 
Determination of the SVD subspace takes about 567 min, in a reasonable agreement with rough 
estimations made in Supporting Information. The SVD-CCSDT calculations for $N_{\mathrm{SVD}}=120$ 
take 
1943 min (22 iterations). Therefore, the total computational cost of the SVD-CCSDT method is about 
five times that of CCSD for this molecule. Finally, the evaluation of the SVD-CCSDT+ correction 
takes ca. 815 min; by comparison, computation of the standard (T) correction takes about 300 min. 
Despite this overhead is non-negligible, we believe that this is a reasonable price to pay for a 
sizable reduction of the error in the final results. All timings reported in this paragraph were 
obtained using a single core of AMD Opteron\textsuperscript{TM} 6174 processor without 
parallelization of the program execution and without exploitation of the point group symmetry.

\section{Conclusions and future work}
\label{sec:conclusion}

In this paper we have reported two novel developments in the field of the rank-reduced CCSDT 
theory. First, we have introduced a non-iterative energy correction, abbreviated as SVD-CCSDT+, 
added on top of the converged 
SVD-CCSDT result in order to approximately account for the triple excitations excluded from the 
parent SVD subspace. The working formula for the correction has been derived by extending the 
Lagrangian formalism of Eriksen \emph{et al.}~\cite{eriksen14b} with an additional assumption that 
the SVD subspace is 
closed under the action of the Fock operator. We have shown that in the limit of complete SVD 
subspace the value of the correction is rigorously equal to zero. In the opposite case of an empty 
SVD subspace the formula for the correction reduces to the well-known (T) method if 
sub-dominant terms that are at least quadratic in the cluster amplitudes are neglected.

The accuracy and computational efficiency of the proposed SVD-CCSDT+ correction has been 
assessed by 
studying a set of isomerization reactions involving small and medium-sized molecular systems. 
We have concluded that the non-iterative correction can fulfill two separate roles. If an accuracy 
level of a fraction of kJ/mol is sufficient, SVD-CCSDT+ correction significantly reduces the size 
of the SVD subspace that has to be employed in the iterative part of the calculations. 
Simultaneously, 
by adding the SVD-CCSDT+ correction the error due to the incompleteness of the SVD subspace can be 
reduced to levels considerably below 0.1 kJ/mol if the SVD subspace size is large enough. This 
levels of accuracy are usually impossible to achieve in practice solely with the SVD-CCSDT method. 
We have also presented representative timings of the SVD-CCSDT and SVD-CCSDT+ calculations, proving 
that the exact CCSDT results can be reproduced to within 0.1 kJ/mol with the computational cost 
only several times larger than required for the CCSD method. The SVD-CCSDT+ method retains 
black-box features of single-reference CC; the size of the SVD subspace 
remains the only additional parameter that has to be specified.

The second theoretical development introduced in this work is an algorithm for 
determination of the triples excitations subspace which is an alternative to the scheme given in 
Ref.~\onlinecite{lesiuk19}. While the formalism proposed here is less general than the 
bidiagonalization strategy from Ref.~\onlinecite{lesiuk19}, it scales rigorously as $N^6$, rather 
than $N^7$, with the system size. Therefore, despite a larger prefactor the proposed method is 
advantageous in applications to larger systems. Moreover, the new method comprises no iterative 
steps which eliminates accumulation of numerical noise and convergence problems one may encounter in 
iterative schemes.

As a final note, in the present work it has been demonstrated that the rank-reduced SVD-CCSDT+ 
method can reliably reproduce the exact CCSDT energetics with significantly decreased computational 
cost. The next important step is incorporation of quadruple excitation effects which become 
important at the 0.1 kJ/mol accuracy level. An economical way to take them into account is offered 
by 
non-iterative schemes such as CCSDT[Q]~\cite{kucharski89,kucharski98,kucharski01} or 
CCSDT(Q)~\cite{bomble05,kallay05,kallay08}. In particular, the latter method was found to 
systematically improve the quality of the results~\cite{eriksen15}, in 
comparison to both CCSD(T) and CCSDT, at a reasonable computational cost. 
Nowadays the CCSDT(Q) method is often regarded as the ''platinum standard`` of the electronic 
structure theory~\cite{kodrycka19}, by analogy to the ''gold standard`` CCSD(T), and is a member of 
composite schemes routinely applied, e.g., in \emph{ab initio} computational 
thermochemistry~\cite{martin99,boese04,karton06,tajti04,bomble06,harding08,feller08}. 
Unfortunately, incorporation of the (Q) correction in the present rank-reduced framework is not 
straightforward as its computational costs scale as $N^9$ with the 
system size if no approximations are introduced. This leads to a question whether by extending the 
rank-reduced CC 
formalism to the quadruply excited amplitudes one can reduce the scaling of the (Q) correction to a 
more manageable $N^7$ level, on par with the SVD-CCSDT+ theory. The answer to this question is 
affirmative and details of the 
procedure will be reported in a separate publication.

\begin{acknowledgement}
I would like to thank M. M\"{o}rchen, Dr. A. Tucholska and Prof. B. Jeziorski for fruitful discussions, and for reading 
and commenting on the manuscript. I am grateful to Prof.~M.~Reiher and all members of his group for their hospitality 
during my stay at Laboratorium f\"{u}r Physikalische Chemie, ETH Z\"{u}rich. This work was supported by the Foundation 
for Polish Science (FNP) and by the Polish National Agency of Academic Exchange through the Bekker programme No. 
PPN/BEK/2019/1/00315/U/00001. Computations presented in this research were carried out with the support of the 
Interdisciplinary Center for Mathematical and Computational Modeling (ICM) at the University of Warsaw, grant number 
G86-1021. 
\end{acknowledgement}

\begin{suppinfo}
The following file is available free of charge via the Internet at \texttt{http://pubs.acs.org}:
\begin{itemize}
  \item {\tt plus-supp.pdf}: derivation of the non-iterative method for determination of the SVD 
  subspace described in Sec. \ref{subsec:pre}, numerical verification of the scaling of the SVD-CCSDT+ method,
  additional numerical results confirming the conclusions of Sec. \ref{subsec:gbctest}, Cartesian coordinates of 
  molecular geometries used in Sec. \ref{subsec:plustot}, and statistical error measures for results presented in Sec. 
  \ref{subsec:plustot}.
\end{itemize}

\end{suppinfo}

\bibliography{svd_plus}

\end{document}


\renewcommand{\arraystretch}{1.25}

\tableofcontents

\newpage

\section{Non-iterative determination of the SVD subspace}
\label{sec:nonit}

In this document we introduce an alternative method of obtaining the SVD subspace of triple 
excitations. In contrast to the algorithm introduced in Ref.~\onlinecite{lesiuk19}, the new method 
is a one-step non-iterative procedure that has a rigorous $N^6$ scaling with the system size. 
However, it is applicable at present only to the approximate triples amplitudes given by Eq. (5) in 
the main paper.

In general, for any approximate $t_{ijk}^{abc}$ amplitudes tensor, the SVD subspace can be found 
by first forming the following symmetric square matrix
\begin{align}
\label{laibj}
 \mathcal{X}_{ai,bj} = t_{ikl}^{acd}\,t_{jkl}^{bcd},
\end{align}
and finding its eigenvectors corresponding to the largest eigenvalues\cite{hino04}. This is a 
consequence of the 
fact that for any rectangular matrix $\mathbf{M}$, its left singular-vectors coincide with the 
eigenvectors of the normal matrix $\mathbf{M}\mathbf{M}^{\mathrm{T}}$. While the diagonalization of 
the $\mathcal{X}_{ai,bj}$ matrix scales as $O^3V^3$ and hence can be obtained efficiently, the cost 
of assembling the matrix $\mathcal{X}_{ai,bj}$ is proportional to $O^4V^4$ if no additional 
simplifications are made. However, we show that for the $\,^{(2)}t_{ijk}^{abc}$ amplitudes, defined 
by Eq. (5) in the main paper, the cost of this step can be reduced down to the level of $N^6$ (from 
now on, we drop the superscript $(2)$ from $\,^{(2)}t_{ijk}^{abc}$ for brevity sake). To this end, 
one must first eliminate the three-particle energy denominator $\epsilon_{ijk}^{abc}$ that 
effectively prevents any factorization of the formula (\ref{laibj}). For this purpose we employ the 
discrete Laplace transformation (LT)
\begin{align}
\label{ltd3}
 (\epsilon_{ijk}^{abc})^{-1} = \sum_g^{N_g} w_g\,e^{-t_g \left( \epsilon_i^a + \epsilon_j^b + 
 \epsilon_k^c \right)},
\end{align}
where $\epsilon_i^a = \epsilon_i - \epsilon_a$, $t_g$ and $w_g$ are the quadrature nodes and 
weights, respectively, and $N_g$ is the quadrature size. This technique was originally introduced 
by 
Alml\"{o}f to get rid of the two-particle energy denominator in the MP2 theory~\cite{almlof91}, but 
it was subsequently employed in other electronic structure 
methods~\cite{haser92,ayala99,lambert05,nakajima06,jung04,kats08}, including CCSD(T)~\cite{pere00}. 
The problem of optimal choice of the quadrature nodes and weights has also been extensively 
discussed in the literature; for the purposes of this work we employ the min-max quadrature 
proposed by Takatsuka \emph{et al}~\cite{takatsuka08,braess05,paris16}. The number of quadrature 
points in Eq. (\ref{ltd3}) is independent of the system size, i.e. $N_g\propto N^0$.

In a straightforward application of the LT technique the formula (\ref{ltd3}) has to be inserted 
twice into Eq. (\ref{laibj}). This leads to the quadratic dependence of the computational costs on 
the number of quadrature points. While this would not change the ultimate scaling of the method it 
would significantly increase the prefactor. To avoid this problem we recall an elementary formula
\begin{align}
\label{ldt3simp}
 \frac{1}{\epsilon_{ikl}^{acd}} \frac{1}{\epsilon_{jkl}^{bcd}} = 
 \frac{1}{\epsilon_j^b-\epsilon_i^a}
 \left( \frac{1}{\epsilon_{ikl}^{acd}} - \frac{1}{\epsilon_{jkl}^{bcd}} \right),
\end{align}
where we have temporarily assumed that $\epsilon_i^a\neq \epsilon_j^b$. Generalization to the 
degenerate case $\epsilon_i^a= \epsilon_j^b$ shall be presented shortly. The 
advantage of this formula is that the product of two three-particle denominators is no longer 
present and a single application of the LT technique is sufficient. By rewriting the triple 
amplitudes as $t_{ijk}^{abc} = \left(\epsilon_{ijk}^{abc}\right)^{-1}\Gamma_{ijk}^{abc}$, inserting 
Eqs. (\ref{ltd3}) and (\ref{ldt3simp}), and after some rearrangements, Eq. (\ref{laibj}) assumes 
the form
\begin{align}
\label{laibj2}
 \mathcal{X}_{ai,bj} = \sum_g^{N_g} w_g\,\frac{e^{-t_g \epsilon_i^a} - e^{-t_g \epsilon_j^b} 
 }{\epsilon_j^b-\epsilon_i^a}\, \Gamma_{ikl}^{acd}\,\Gamma_{jkl}^{bcd}\,
 e^{-t_g \left( \epsilon_k^c + \epsilon_l^d\right) },\;\;\;\mbox{for}\;\;
 \epsilon_j^b\neq\epsilon_i^a.
\end{align}
The case of degenerate orbital energy differences can now be resolved by analytically taking the 
limit $\epsilon_j^b\rightarrow\epsilon_i^a$. This leads to a separate expression for the 
remaining elements
\begin{align}
\label{laibj3}
 \mathcal{X}_{ai,bj} = \sum_g^{N_g} w_g\,t_g\,e^{-t_g \epsilon_i^a}\,
 \Gamma_{ikl}^{acd}\,\Gamma_{jkl}^{bcd}\,
 e^{-t_g \left( \epsilon_k^c + \epsilon_l^d\right) },\;\;\;\mbox{for}\;\;
 \epsilon_j^b=\epsilon_i^a.
\end{align}

The formulas (\ref{laibj2}) and (\ref{laibj3}) show that the matrix $\mathcal{X}_{ai,bj}$ can be 
obtained as a weighted sum of simpler quantities in the form
\begin{align}
\label{small_l}
 x_{ai,bj} = \Gamma_{ikl}^{acd}\,\Gamma_{jkl}^{bcd}\,
 e^{-t_g \left( \epsilon_k^c + \epsilon_l^d\right)}
\end{align}
where the three-particle denominators are no longer present enabling an efficient 
factorization. To improve the efficiency of the calculations we additionally employ the 
following decomposition of the doubly excited CCSD amplitudes
\begin{align}
\label{t2diag}
 t_{ij}^{ab} = V_{ai}^A\,s_A\,V_{bj}^A,
\end{align}
which is obtained by diagonalization treating $t_{ij}^{ab}$ as a symmetric $OV\times OV$ matrix.
The advantage of this formula is 
that the eigenvectors corresponding to eigenvalues below a certain threshold, i.e., 
$|s_A|<\epsilon$, can be dropped. As demonstrated by Parrish \emph{et al.}~\cite{parrish19}, 
for a 
fixed value of $\epsilon>0$ the number of retained eigenvalues, denoted $N_{\mathrm{eig}}$ further 
in the text, scales linearly with the system size. The cost of computing the eigendecomposition is 
proportional to $O^3V^3$ which is acceptable from the present point of view. If necessary, this 
cost can be reduced with the help of an 
iterative Lanczos-type diagonalization method which targets only the dominant 
eigenpairs. 
However, this possibility has not been explored in the present work.

Consider the following fixed form of the $\Gamma_{ijk}^{abc}$ tensor
\begin{align}
\label{gammaijk}
 \Gamma_{ijk}^{abc} = \langle \,_{ijk}^{abc} | \big[ W, T_2 \big] \rangle = \left( 1 + P_{bj,ck} 
\right) \left( 1 + P_{ai,bj} + P_{ai,ck}\right)
 \Big[ t_{ij}^{ad}\,(ck|bd) - t_{il}^{ab}\,(ck|lj) \Big],
\end{align}
which corresponds to the triples amplitudes obtained in the leading order of the perturbation 
theory. For simplicity, in the above formula we employ the standard two-electron integrals 
corresponding to the pure fluctuation potential $W$. However, exactly the same working formulas 
hold for the $T_1$-transformed Hamiltonian, $\widetilde{W}=e^{-T_1}We^{T_1}$. To cover this case it 
is sufficient to replace the conventional 
two-electron integrals $(pq|rs)$ by their $T_1$-transformed counterparts, $(pq\widetilde{|}rs)$, 
see Sec. 2.1 in the main paper. To facilitate an efficient evaluation of Eq.
(\ref{small_l}) with $\Gamma_{ijk}^{abc}$ given by Eq. (\ref{gammaijk}) we define the following 
intermediate that is independent of the Laplace grid points
\begin{align}
 B_{ai}^{QA} = B_{ae}^Q\,V_{ei}^A\,s_A - B_{mi}^Q\,V_{am}^A\,s_A,
\end{align}
and a handful of grid-dependent intermediates
\begin{align}
 &Z_{PQ} = B_{ai}^P\,B_{ai}^Q\,e^{-t_g \epsilon_i^a},\;\;
 Y_A^P  = B_{ai}^P\,V_{ai}^A\,e^{-t_g \epsilon_i^a}, \;\;
 W_{AB} = V_{ai}^A\,V_{ai}^B\,e^{-t_g \epsilon_i^a}, \\
 &I_A^{PQ} = B_{ai}^{QA}\,B_{ai}^P\,e^{-t_g \epsilon_i^a},\;\;
 J_{AB}^P = B_{ai}^{PA}\,V_{ai}^B\,e^{-t_g \epsilon_i^a}.
\end{align}
Calculation of the above intermediates scales as $\propto N^5$ or lower and it does not 
constitute a bottleneck. With help of these quantities $x_{ai,bj}$ can be 
rewritten as a sum of only six topologically distinct terms
\begin{align}
 x_{ai,bj} = x_{ai,bj}^{(1)} + x_{ai,bj}^{(4)} + x_{ai,bj}^{(6)}
 + \left( 1 + P_{ai,bj} \right) \Big[ x_{ai,bj}^{(2)} + x_{ai,bj}^{(3)} + x_{ai,bj}^{(5)} \Big],
\end{align}
where
\begin{align}
 &x_{ai,bj}^{(1)} = V_{ai}^A\,V_{bj}^B\,\Big[ 
 B_{ck}^{QA}\,e^{-t_g \epsilon_k^c}\,B_{ck}^{PB}\,Z_{PQ} + I_A^{QP}\,I_B^{PQ}
 \Big], \\
 &x_{ai,bj}^{(2)} = V_{ai}^B B_{bj}^{QA} \Big[ Z_{PQ}\,J_{BA}^P + Y_A^P\,I_B^{PQ} \Big], \\
 &x_{ai,bj}^{(3)} = V_{ai}^B\,B_{bj}^Q\,\Big[ B_{ck}^{PB} e^{-t_g \epsilon_k^c}\,
 B_{ck}^{QA}\,C_A^P + I_A^{QP} J_P^{BA} \Big],\\
 &x_{ai,bj}^{(4)} = B_{ai}^{PB}\,B_{bj}^{QA}\Big[Z_{PQ} W_{AB} + C_A^P\,C_B^Q \Big],\\
 &x_{ai,bj}^{(5)} = B_{ai}^{QA}\,B_{bj}^P\Big[ C_B^Q\,J_{BA}^P + I_B^{PQ}\,W_{AB} \Big],\\
 &x_{ai,bj}^{(6)} = B_{ai}^Q\,B_{bj}^P \Big[
 B_{ck}^{QA}\,B_{ck}^{PB}\,e^{-t_g \epsilon_k^c}\,W_{AB} + J_{AB}^Q\,J_{BA}^P \Big].
\end{align}
The computational cost of the whole procedure is dominated by the fourth term, $x_{ai,bj}^{(4)}$, 
where 
the rate-determining step scales as $O^2V^2N_{\mathrm{eig}}N_{\mathrm{aux}}$. Let us compare this 
with the most expensive term in the CCSD equations, the so-called particle-particle ladder 
diagram, with the scaling proportional to $O^2V^4$. As demonstrated in the next section, in 
practice it is 
reasonable to assume that $N_{\mathrm{eig}}\approx V$. This means that the cost of evaluating 
$x_{ai,bj}^{(4)}$ is more expensive than a single CCSD iteration by a factor of 
$N_{\mathrm{aux}}/V\approx 2-4$. To this one has to add that $x_{ai,bj}^{(4)}$ must be evaluated at 
each Laplace 
grid point and $3-5$ points are usually required for a sufficient accuracy. Therefore, the cost of 
assembling the complete $\mathcal{X}_{ai,bj}$ matrix is expected to be comparable to $5-15$ 
CCSD iterations, the exact ratio depending on the desired accuracy level. While this prefactor is 
substantial, the main advantage of the proposed method is that the scaling of this procedure is the 
same as of the CCSD theory and this ratio is guaranteed to approximately hold for arbitrarily 
large systems.

\newpage

\section{Parameters in the non-iterative SVD subspace determination}
\label{sec:svdtest}

\begin{table}[b!h]
  \caption{Relative error (in parts per million) in the SVD-CCSDT correlation energy for the 
methane molecule (cc-pVTZ basis set) as a function of the eigenvalue threshold ($\epsilon$) and 
number of Laplace grid points ($N_g$). The reference results were generated with $\epsilon=0$ 
and $N_g=20$.}
  \label{tab:nonit1}
  \begin{tabular}{rrrrrrr}
    \hline\hline
    \multicolumn{1}{c}{$\epsilon$} & \multicolumn{6}{c}{$N_g$} \\
    \hline
      & \multicolumn{1}{c}{2} 
      & \multicolumn{1}{c}{3} 
      & \multicolumn{1}{c}{4} 
      & \multicolumn{1}{c}{5} 
      & \multicolumn{1}{c}{7} 
      & \multicolumn{1}{c}{10} 
      \\
    \hline
     \multicolumn{7}{c}{$N_{\mathrm{SVD}}=25$} \\
    \hline
    $10^{-2}$ & 1314 & 1417 & 1425 & 1423 & 1423 & 1423 \\
    $10^{-3}$ & 469  & 250  & 214  & 219  & 212  & 221 \\
    $10^{-4}$ & 474  & 251  & 216  & 221  & 224  & 223 \\
    $10^{-5}$ & 602  & 139  & 51   & 38   & 18   & <0 \\
    \hline
     \multicolumn{7}{c}{$N_{\mathrm{SVD}}=50$} \\
    \hline
    $10^{-2}$ & 360 & 138 & 111 & 96 & 101 & 101 \\
    $10^{-3}$ & 389 & 113 & 101 & 87 & 87  & 89 \\
    $10^{-4}$ & 389 & 111 & 66  & 85 & 85  & 87 \\
    $10^{-5}$ & 312 & 36  & 17  & 2  & <0  & <0 \\
    \hline
     \multicolumn{7}{c}{$N_{\mathrm{SVD}}=75$} \\
    \hline
    $10^{-2}$ & 4   & 307 & 267 & 240 & 252 & 252 \\
    $10^{-3}$ & 202 & 52  & 6   & 54  & 66  & 66 \\
    $10^{-4}$ & 201 & 59  & 12  & 54  & 66  & 66 \\
    $10^{-5}$ & 234 & 49  & 13  & 7   &  1  & <0 \\
    \hline\hline
  \end{tabular}
\end{table}

The method of non-iterative determination of the SVD subspace introduced in here
depends on two parameters that control the accuracy. The first one, denoted $\epsilon$ further in 
the text, is the truncation threshold in Eq. (\ref{t2diag}). Eigenvalue-eigenvector pairs with 
$|s_A|<\epsilon$ are dropped from the expansion (\ref{t2diag}), reducing its length to
$N_{\mathrm{eig}}$. The second parameter ($N_g$) is the number of quadrature points in the 
Laplace transformation of the three-particle energy denominator, Eq. (\ref{ltd3}). In this section 
we determine values of these parameters which minimize the computational effort necessary 
to find the SVD subspace, but simultaneously guarantee a satisfactory accuracy level in the 
final results.

\begin{table}[t]
  \caption{The same data as in Table \ref{tab:nonit1} but for the formaldehyde molecule (cc-pVTZ 
basis set).}
  \label{tab:nonit2}
  \begin{tabular}{rrrrrrr}
    \hline\hline
    \multicolumn{1}{c}{$\epsilon$} & \multicolumn{6}{c}{$N_g$} \\
    \hline
      & \multicolumn{1}{c}{2} 
      & \multicolumn{1}{c}{3} 
      & \multicolumn{1}{c}{4} 
      & \multicolumn{1}{c}{5} 
      & \multicolumn{1}{c}{7} 
      & \multicolumn{1}{c}{10} 
      \\
    \hline
     \multicolumn{7}{c}{$N_{\mathrm{SVD}}=50$} \\
    \hline
    $10^{-2}$ & 251 & 426 & 450 & 447 & 450 & 451 \\
    $10^{-3}$ & 54  & 178 & 20  & 32  & 30  & 29 \\
    $10^{-4}$ & 18  & 139 & 23  & 5   & 1   & 14 \\
    $10^{-5}$ & 17  & 138 & 23  & 7   & 6   & <0 \\
    \hline
     \multicolumn{7}{c}{$N_{\mathrm{SVD}}=100$} \\
    \hline
    $10^{-2}$ & 202 & 287 & 243 & 263 & 259 & 258 \\
    $10^{-3}$ & 19  & 33  & 11  & 29  & 27  & 26 \\
    $10^{-4}$ & 32  & 45  & 4   & 7   & 2   & 1 \\
    $10^{-5}$ & 33  & 45  & 5   & 10  & 4   & <0 \\
    \hline
     \multicolumn{7}{c}{$N_{\mathrm{SVD}}=150$} \\
    \hline
    $10^{-2}$ & 4   & 122 & 206 & 208 & 205 & 202 \\
    $10^{-3}$ & 203 & 69  & 7   & 2   & 7   & 4 \\
    $10^{-4}$ & 201 & 64  & 3   & <0  & 3   & 1 \\
    $10^{-5}$ & 202 & 66  & 2   & <0  & <0  & <0 \\
    \hline\hline
  \end{tabular}
\end{table}

To this end we carried out SVD-CCSDT calculations for a dozen or so 
polyatomic molecules. The SVD vectors were determined using the method described in the previous 
section with different values of the control parameters $\epsilon$ and $N_g$. For 
several fixed $N_{\mathrm{SVD}}$ we then recorded the error in the SVD-CCSDT correlation energy as 
a function of $\epsilon$ and $N_g$. As reference values we employ SVD-CCSDT results obtained 
with $\epsilon=0$, i.e. without truncation of the expansion (\ref{t2diag}), and for $N_g=20$ 
Laplace quadrature points. This is sufficient to make the reference data essentially exact for the 
present purposes.

We present results for two representative systems: methane molecule 
(tetrahedral geometry with $R_{\mathrm{C-H}}=1.0838\,$\AA{}) and formaldehyde molecule (planar 
$C_{2v}$ geometry with $R_{\mathrm{C-H}}=1.1005\,$\AA{}, $R_{\mathrm{C-O}}=1.2105\,$\AA{}, and 
$\theta_{\mathrm{H-C-O}}=121.92^\circ$). The results are presented in Tables \ref{tab:nonit1} and 
\ref{tab:nonit2}, respectively. For both molecules we provide results for three different SVD 
subspace dimensions, purposefully chosen to be small, moderate and large from the point of view of 
realistic applications. In the case of the large SVD subspace we demand that the error in the 
correlation energy resulting from approximations in Eqs. (\ref{t2diag}) and (\ref{ltd3}) does not 
exceed 100 parts per million (0.01\%). We expect this level of accuracy to be sufficient in most 
applications. For small SVD subspaces the accuracy requirement can be less stringent and 
an error of about one part per thousand (0.1\%) is acceptable. 

By inspection 
of the results given in Tables \ref{tab:nonit1} and \ref{tab:nonit2} one concludes that the 
combination of the eigenvalue threshold $\epsilon=10^{-3}$ with $N_g=4$ Laplace grid points is 
sufficient to reach the prescribed accuracy levels in all cases. While in some examples less strict 
values of the control parameters give unexpectedly small errors, this is probably  
accidental and is not expected to hold for a broader range of systems. 
Therefore, we recommend to set $\epsilon=10^{-3}$ and $N_g=4$ as a default in the 
algorithm described above. These values of the control parameters were 
employed in all calculations reported in this work.

\newpage

\section{Numerical verification of the computational complexity}
\label{sec:scaling}

\begin{figure}[h]
\includegraphics[scale=0.75]{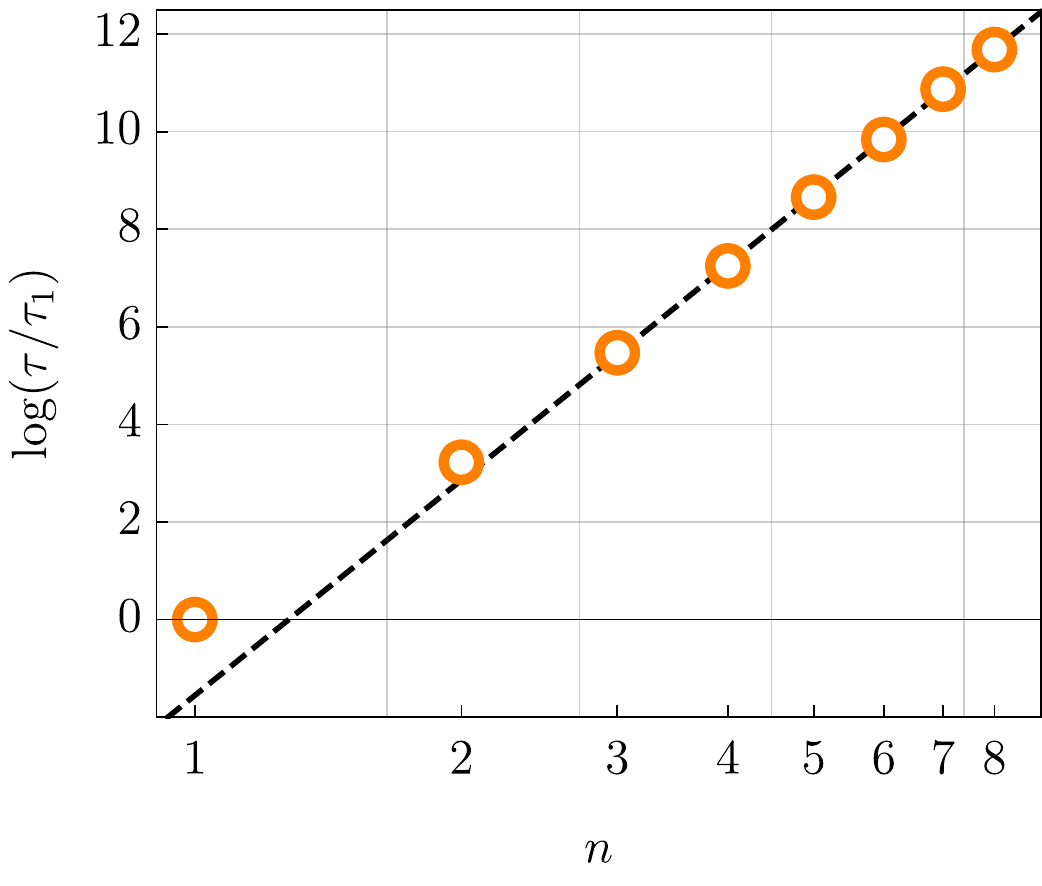}
\caption{\label{timings} Timings $\tau$ of the $\delta E_{\rm T+}$ calculations for linear alkanes C$_n$H$_{2n+2}$ as a 
function of the chain length, $n$. The timings are given relative to the calculations for $n=1$ (denoted by the symbol 
$\tau_1$) and the plot employs the doubly-logarithmic scale. The dashed black curve is a linear function 
($6.37x-1.55$) fitted to the data points for $n=3,\ldots,8$.
}
\end{figure}

To study the computational complexity of evaluating the $\delta E_{\rm T+}$ correction we carried out 
the calculations for linear alkanes C$_n$H$_{2n+2}$, where $n=1,2,\ldots,8$ denotes the chain length. The geometries 
of all molecules were taken from Ref.~\onlinecite{lesiuk20}. We employ the cc-pVDZ basis set together with the 
corresponding cc-pVDZ-RI density-fitting basis. For each molecule the SVD subspace size is equal to the number of 
virtual orbitals in the system ($N_{\mathrm{SVD}}=V$), according to the recommendations from the main text. The 
computational timings are reported in Fig. \ref{timings}.

\newpage

\section{Further numerical verification of the condition (14)}
\label{sec:gbctest2}

\begin{figure}[h]
\includegraphics[scale=0.75]{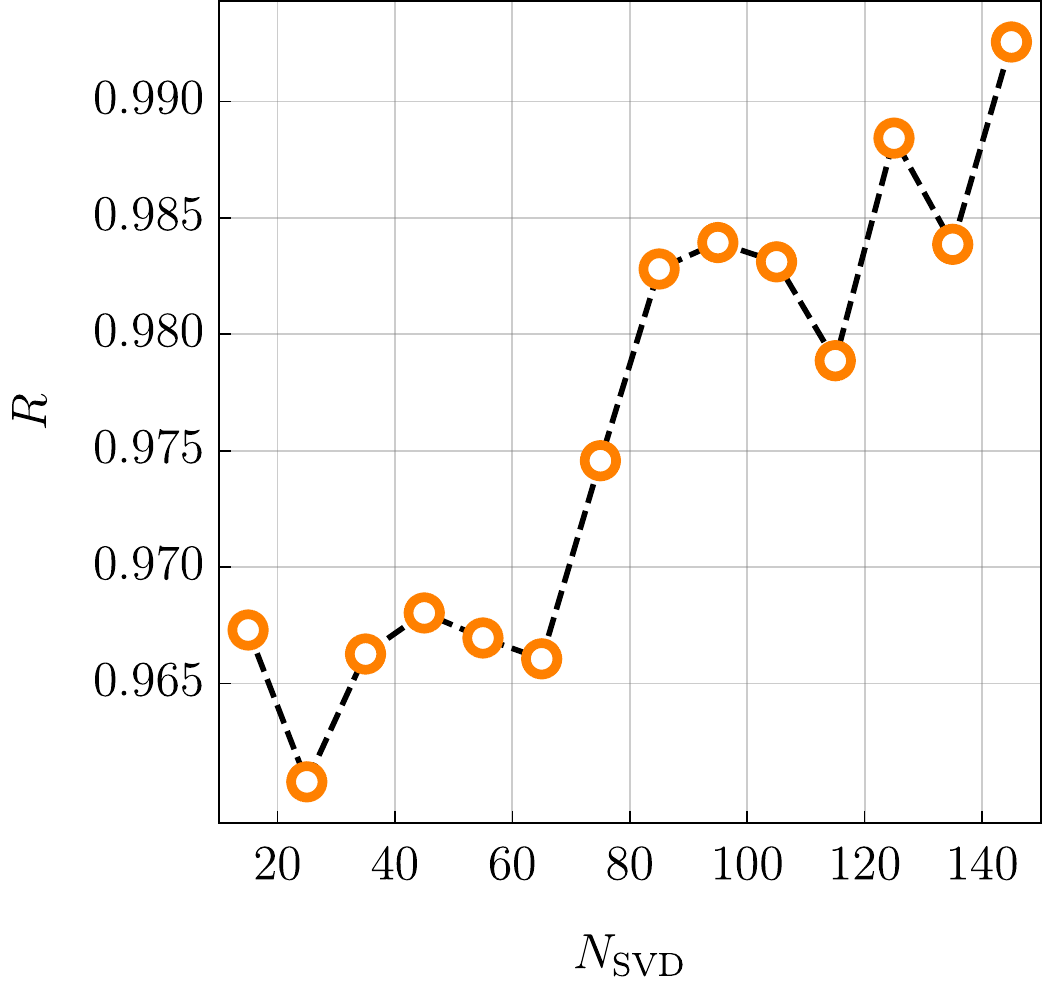}
\caption{\label{gbc_h2o} Values of the $R$ coefficient, see Eq. (40) in the main paper, as a 
function of the SVD subspace size for the H$_2$O molecule ($C_{2v}$ symmetry group, 
$R_{\mathrm{O-H}}=0.9562\,$\AA{}, $R_{\mathrm{H-H}}=1.5096\,$\AA{}, cc-pVTZ basis set, $1s^2$ 
oxygen core orbital frozen). The black dashed lines are linear functions connecting two neighboring 
data points. The maximum size of the SVD subspace for this system equals to $212$.}
\end{figure}

\begin{figure}
\includegraphics[scale=0.75]{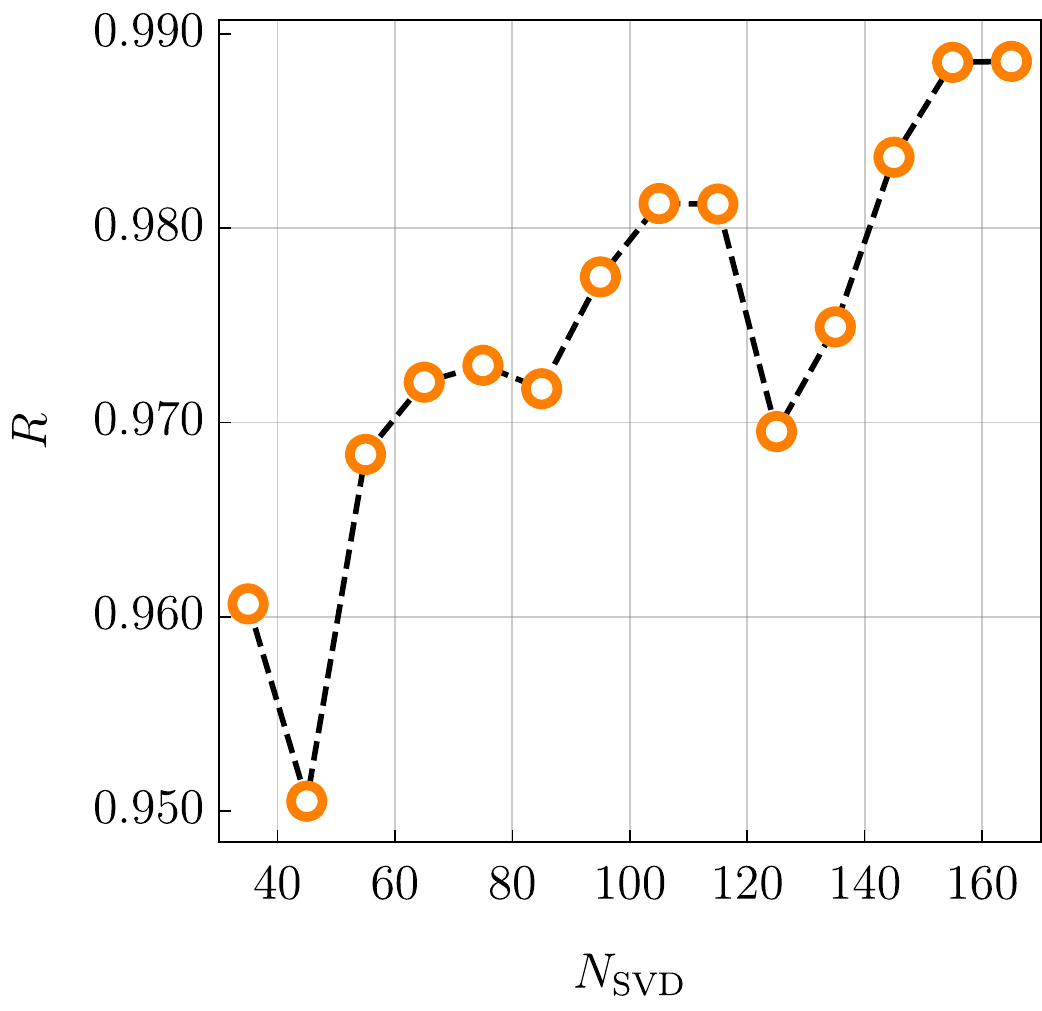}
\caption{\label{gbc_nh3} The same data as in Fig. \ref{gbc_h2o}, but for the NH$_3$ molecule 
($C_{3v}$ symmetry group, $R_{\mathrm{N-H}}=1.0101\,$\AA{}, $R_{\mathrm{H-H}}=1.6173\,$\AA{}, 
cc-pVTZ basis set, $1s^2$ nitrogen core orbital frozen). The maximum size of the SVD subspace for 
this system equals to $268$.}
\end{figure}

\begin{figure}
\includegraphics[scale=0.75]{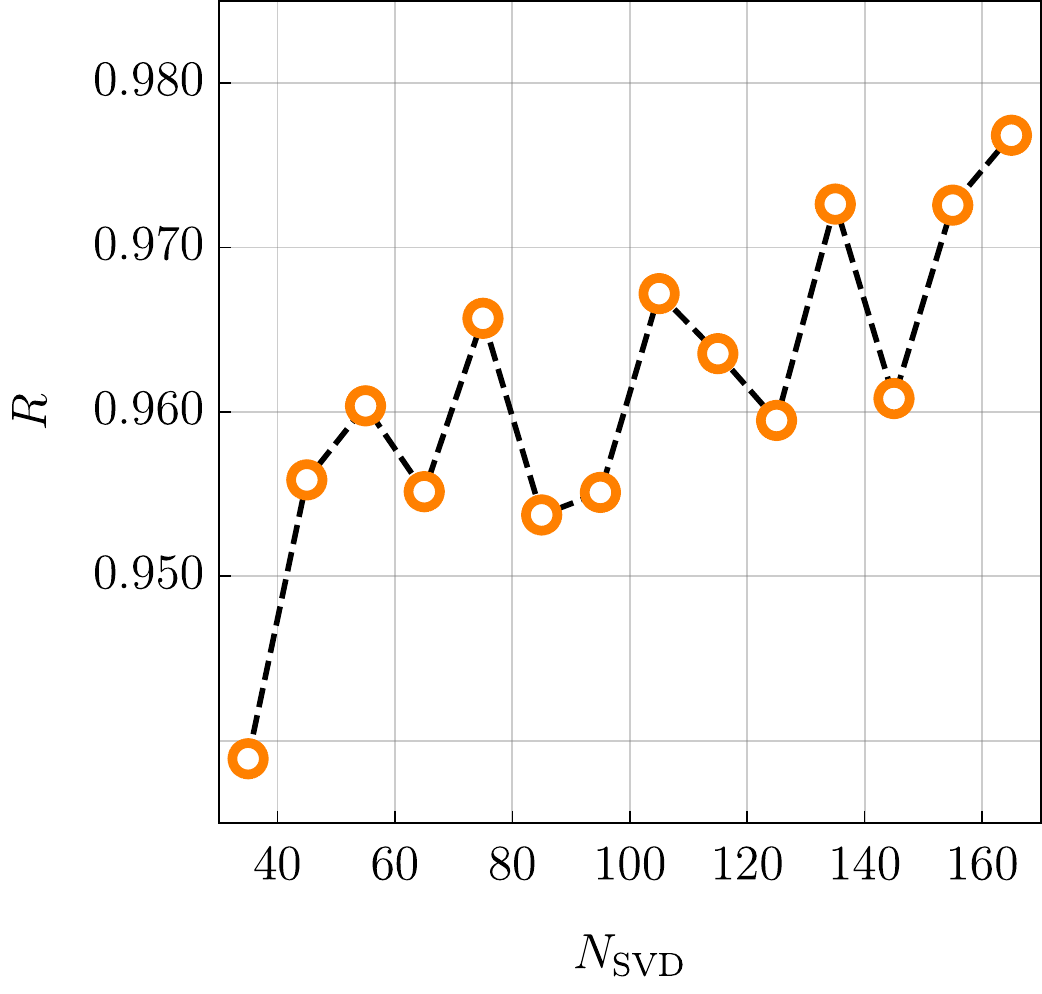}
\caption{\label{gbc_hcch} The same data as in Fig. \ref{gbc_h2o}, but for the HCCH molecule 
(linear geometry, $R_{\mathrm{C-H}}=1.0613\,$\AA{}, $R_{\mathrm{C-C}}=1.2041\,$\AA{}, cc-pVTZ basis set, 
two $1s^2$ carbon core orbitals frozen). The maximum size of the SVD subspace for this system 
equals to $486$.}
\end{figure}

\newpage

\section{Cartesian coordinates of molecular geometries used \\ in Sec. 3.3 of the main paper}
\label{sec:geom}

All coordinates are given in \AA{}ngstr\"{o}ms.

\setlength{\tabcolsep}{20pt}
\begin{table}[H]
  \label{tab:geom_xyz_1}
  \begin{tabular}{rrrr}
  \underline{BH$_3$} & & & \\
  B &    0.0000 &    0.0000 & 0.0000 \\
  H &    0.0000 &    1.1900 & 0.0000 \\
  H &    1.0306 & $-$0.5950 & 0.0000 \\
  H & $-$1.0306 & $-$0.5950 & 0.0000 \\
  \underline{C$_2$H$_2$} & & & \\
  H & 0.0000 & 0.0000 & $-$1.0613 \\
  C & 0.0000 & 0.0000 &    0.0000 \\
  C & 0.0000 & 0.0000 &    1.2041 \\
  H & 0.0000 & 0.0000 &    2.2654 \\
  \underline{CH$_4$} & & & \\
  C &    0.0000 &    0.0000 &    0.0000 \\
  H & $-$0.6271 &    0.6271 &    0.6271 \\
  H &    0.6271 & $-$0.6271 &    0.6271 \\
  H & $-$0.6271 & $-$0.6271 & $-$0.6271 \\
  H &    0.6271 &    0.6271 & $-$0.6271 \\
  \underline{CO$_2$} & & & \\
  O & 0.0000 & 0.0000 & $-$1.1601 \\
  C & 0.0000 & 0.0000 &    0.0000 \\
  O & 0.0000 & 0.0000 & $+$1.1601 \\
  \underline{CO} & & & \\
  C & 0.0000 & 0.0000 & $-$0.6508 \\
  O & 0.0000 & 0.0000 &    0.4882 \\
  \underline{NH$_3$} & & & \\
  N &    0.0000 & 0.0000 &    0.0000 \\
  H &    1.0101 & 0.0000 &    0.0000 \\
  H & $-$0.2845 & 0.0000 & $-$0.9692 \\
  H & $-$0.2845 & 0.8394 &    0.4846 \\
  \end{tabular}
\end{table}

\begin{table}[H]
  \label{tab:geom_xyz_2}
  \begin{tabular}{rrrr}
  \underline{F$_2$} & & & \\
  F & 0.0000 & 0.0000 & 0.0000 \\
  F & 0.0000 & 0.0000 & 1.4111 \\
  \underline{H$_2$CO} & & & \\
  C &     0.0000 & 0.0000 & $-$0.6204 \\
  O &     0.0000 & 0.0000 &    0.5925 \\
  H &     0.9360 & 0.0000 & $-$1.1986 \\
  H &  $-$0.9360 & 0.0000 & $-$1.1986 \\
  \underline{H$_2$O$_2$} & & & \\
  O &    0.0000 & 0.0000 &    0.0000 \\
  O &    1.4495 & 0.0000 &    0.0000 \\
  H & $-$0.1669 & 0.0000 & $-$0.9464 \\
  H &    1.6164 & 0.8747 &    0.3612 \\
  \underline{H$_2$O} & & & \\
  O &    0.0000 & 0.0000 &    0.0000 \\
  H &    0.9562 & 0.0000 &    0.0000 \\
  H & $-$0.2354 & 0.0000 & $-$0.9268 \\
  \underline{HCN} & & & \\
  N & 0.0000 & 0.0000 &    0.5944 \\
  C & 0.0000 & 0.0000 & $-$0.5726 \\
  H & 0.0000 & 0.0000 & $-$1.6371 \\
  \underline{HCNO} & & & \\
  O &    1.1624 & $-$0.0095 & 0.0000 \\
  C & $-$0.0104 & $-$0.0420 & 0.0000 \\
  N & $-$1.2232 &    0.0984 & 0.0000 \\
  H & $-$1.8750 & $-$0.6671 & 0.0000 \\
  \underline{HF} & & & \\
  H & 0.0000 & 0.0000 & 0.0000 \\
  F & 0.0000 & 0.0000 & 0.9152 \\
  \underline{N$_2$} & & & \\
  N & 0.0000 & 0.0000 & 0.0000 \\
  N & 0.0000 & 0.0000 & 1.0981 \\
  \end{tabular}
\end{table}

\begin{table}[H]
  \label{tab:geom_xyz_3}
  \begin{tabular}{rrrr}
  \underline{HCO$_2$H} & & & \\
  C &    0.4159 &    0.0922 & 0.0000 \\
  O & $-$0.2086 &    1.1230 & 0.0000 \\
  O & $-$0.1250 & $-$1.1409 & 0.0000 \\
  H & $-$1.0882 & $-$1.0190 & 0.0000 \\
  H &    1.5057 &    0.0155 & 0.0000 \\
  \underline{N$_2$O} & & & \\
  N & $-$1.7980 &    1.5224 & 0.0000 \\
  N & $-$0.9703 &    0.7222 & 0.0000 \\
  O & $-$0.1239 & $-$0.0961 & 0.0000 \\
  \end{tabular}
\end{table}

\newpage

\section{Statistical measures of the relative errors defined \\ in Sec. 3.3 of the main paper}
\label{sec:statistics}

\begin{table}[H]
  \caption{Statistical measures (in percent) of the relative error in the total SVD-CCSDT and 
SVD-CCSDT+ correlation energies with respect to the exact CCSDT method. The quantities 
$\bar{\Delta}$ and $\Delta_{\mathrm{std}}$ are defined in the main text, Eqs. (42) and (43). The 
parameter $x$ defines the size of the triple excitation subspace, $N_{\mathrm{SVD}}=x\cdot 
N_{\mathrm{MO}}$, where $N_{\mathrm{MO}}$ is the total number of orbitals in a given system.}
  \label{tab:statistics}
  \begin{tabular}{crrrr}
    \hline\hline
    $x$ & \multicolumn{2}{c}{SVD-CCSDT} & \multicolumn{2}{c}{SVD-CCSDT+} \\
        & \multicolumn{1}{c}{$\bar{\Delta}$}
        & \multicolumn{1}{c}{$\Delta_{\mathrm{std}}$}
        & \multicolumn{1}{c}{$\bar{\Delta}$}
        & \multicolumn{1}{c}{$\Delta_{\mathrm{std}}$} \\
    \hline
    $\frac{1}{6}$ & $-1.2970$ & $0.5388$ & $-0.1632$ & $0.0432$ \\
    $\frac{1}{3}$ & $-0.5676$ & $0.6710$ & $-0.1020$ & $0.0453$ \\
    $\frac{1}{2}$ & $-0.0855$ & $0.1478$ & $-0.0637$ & $0.0363$ \\
    $\frac{2}{3}$ & $0.0730$  & $0.0936$ & $-0.0491$ & $0.0293$ \\
    $\frac{5}{6}$ & $0.1166$  & $0.0469$ & $-0.0391$ & $0.0245$ \\
    $1$           & $0.1135$  & $0.0476$ & $-0.0321$ & $0.0203$ \\
    \hline\hline
  \end{tabular}
\end{table}

\newpage

\bibliography{svd_plus}